\documentclass[floatfix,aps,prd,showpacs,amsmath,nofootinbib,
preprintnumbers,onecolumn]{revtex4}

\usepackage{colordvi}
\usepackage{graphicx}

\usepackage{dcolumn}
\usepackage{bm}

\usepackage{amsmath,amssymb}

\newcommand{\RI}{\mathbb{R}}
\newcommand{\HI}{\mathbb{H}}

\newcommand{\TI}{\mathbb{T}}

\def\etal{{\it et al}.} \def\e{{\rm e}}
\def\dd{{\rm d}} \def\ds{\dd s}  \def\de{\delta}
 \def\al{\alpha} \def\vph{\varphi}
\def\ka{\kappa} \def\rh{\rho}
\font\sevenrm=cmr7 \def\ns#1{_{\hbox{\sevenrm #1}}}
\def\beq{\begin{equation}}\def\eeq{\end{equation}}
\def\bea{\begin{eqnarray}}\def\eea{\end{eqnarray}}
\def\Z#1{_{\lower2pt\hbox{$\scriptstyle#1$}}}
\def\X#1{_{\lower2pt\hbox{$\scriptscriptstyle#1$}}}
\def\MM#1{{\cal M}^{#1}}
\def\PRL#1{Phys.\ Rev.\ Lett.\ {\bf#1}} \def\PR#1{Phys.\ Rev.\ {\bf#1}}
\def\CQG#1{Class.\ Quantum Grav.\ {\bf#1}}
\def\ApJ#1{Astrophys.\ J.\ {\bf#1}} \def\AsJ#1{Astron.\ J.\ {\bf#1}}

\begin{document}

\parskip 4pt 

\title{Cosmic acceleration from M theory on twisted spaces}

\author{Ishwaree P. Neupane}\email{ishwaree.neupane@canterbury.ac.nz}
\author{David L. Wiltshire}\email{david.wiltshire@canterbury.ac.nz}

\affiliation{Department of Physics and Astronomy, University
of Canterbury, Private Bag 4800, Christchurch, New Zealand}

\begin{abstract}

In a recent paper [I.~P.~Neupane and D.~L.~Wiltshire, Phys. Lett. B 
{\bf 619}, 201 (2005).] we have found a new class of
accelerating cosmologies arising from a time--dependent
compactification of classical supergravity on product spaces that
include one or more geometric twists along with non-trivial curved
internal spaces. With such effects, a scalar potential can have a
local minimum with positive vacuum energy. The existence of such a
minimum generically predicts a period of accelerated expansion in
the four-dimensional Einstein-conformal frame. Here we extend our
knowledge of these cosmological solutions by presenting new
examples and discuss the properties of the solutions in a more
general setting. We also relate the known (asymptotic) solutions
for multi-scalar fields with exponential potentials to the
accelerating solutions arising from simple (or twisted) product
spaces for internal manifolds.

\end{abstract}



\maketitle

\section{Introduction}

The possibility that fundamental scalar fields which are uniform
in space play a preeminent role on cosmological scales has been
confirmed by a decade of observations. Most recently the WMAP
measurements of fine details of the power spectrum of cosmic
microwave background anisotropies \cite{wmap} have lent strong
support to the idea that the universe underwent an early
inflationary expansion at high energy scales. The WMAP data, along
with the independent observations of the dimming of type Ia
supernovae in distant galaxies \cite{Riessetal99} are also usually
interpreted as an indication that the universe is undergoing
accelerated expansion at the present epoch, albeit at a vastly
lower rate.

As in the case of the much earlier period of early universe
inflation, the most natural explanation for the repulsive force
responsible for accelerating cosmologies would be a fundamental
vacuum energy, possibly in the form of one or more dynamical
homogeneous isotropic scalar fields. The nature of this dark
energy, which constitutes of order 70\% of the matter--energy
content of the Universe at the present epoch, constitutes a
mystery whose explanation is possibly the greatest challenge faced
by the current generation of cosmologists.

In the $\Lambda$CDM model, the dark energy at the present epoch is
attributed purely to a constant vacuum energy (or cosmological constant),
or equivalently a homogeneous isotropic fluid whose pressure, $p$, and
energy density, $\rh$, are related by $\rho=-p=\Lambda$. This is only the
simplest (and perhaps most common) explanation.
From a field theoretic viewpoint it would be perhaps more natural
to attribute the dark energy to one or more dynamical scalar fields
\cite{early_quintessence,qreview}. Many such quintessence models have
been studied, with scalar potentials which range from completely
{\it ad hoc} ones to those with various theoretical motivations.
Such motivations are often more than not phenomenological. For
example, ultra light pseudo Nambu--Goldstone bosons can give realistic
cosmologies \cite{pngb}, even if one does not specify exactly what
fundamental theory such scalars belong to.

Fundamental scalar fields are of course abundant in higher--dimensional
theories of gravity. The typical scalar potentials that one obtains
by dimensional reduction, namely exponential potentials have been
widely studied \cite{halliwell} but generally without regard to the
restrictions on the sign of the potential and magnitude
of the coupling constants that arise from compactifications or
the higher--dimensional geometry.
In the case of multiple scalar fields, for example, attention has been
focused on simplified potentials in which each scalar appears only once
~\cite{Ed99a,Lidsey:1992ak,Barreiro:1999zs,Heard:2002dr,Guo:2003rs}.

A number of models with a fundamental higher--dimensional origin,
which accommodate a 4--dimensional universe with accelerating expansion
have been studied over the past two decades \cite{acc_uni1,Litterio,acc_uni2},
usually in relation to inflationary epochs in the very early universe.
A general feature of these models was the presence of additional
degrees of freedom, such as fluxes or a cosmological
term. Until relatively recently it was not believed that one
could obtain accelerating universes in the Einstein frame in four
dimensions by compactification of pure Einstein gravity in higher
dimensions \cite{Levin}. In fact, the result was elevated to the
status of a no--go theorem \cite{nogo1}.

Townsend and Wohlfarth \cite{TW} circumvented the no--go theorem
by relaxing its conditions to obtain time-dependent (cosmological)
compactifications
of pure Einstein gravity in arbitrary dimensions, with negatively
curved internal space, which are made compact by topological
identifications. Their exact solutions exhibit a transient
epoch of acceleration between decelerating epochs at early and late
times, {\it albeit} with the number of e-folds of order unity.
Applied to late--time cosmologies, this could mean that our
present cosmic acceleration is not eternal but switches off in a
natural way.

Transient acceleration was subsequently shown to be a generic feature
of many supergravity compactifications
\cite{CMChen:2002,OhtaPRL03,MNR03a,Emparan03a,%
CHNW03b,CHNOW03b,Roy:2003nd,Wohlfarth:2003kw,Townsend03b,IPN03c,IPN03d},
some models including higher--dimensional fluxes appropriate to particular
supergravity models, and others with more complicated internal product
spaces with negatively curved factors.
The generic nature of transient acceleration
is understood also for some simple potentials arising from group manifold
reductions~\cite{Bergshoeff:2003vb}, and a compactification of M--theory on
a singular Calabi-Yau space~\cite{Jarv:2003qy}.
All these solutions typically exhibit a short period of accelerated
expansion. 
Other approaches to accelerating cosmologies
involve~\cite{Maeda:2004hu} a compactification of string/M theory
with higher order curvature corrections, such as $(\mbox{Weyl})^4$
terms in type II string theory, see also a
review~\cite{Ohta:2004wk}. In this rather complicated scenario,
the analysis is often restricted to asymptotic solutions of the
evolution equations, which involve a fine-tuning of the string
coupling or the Planck scale in the higher dimensions (10 or 11).
It is not immediately clear if any asymptotic solutions would be
available within the full string theories.

In a recent paper \cite{IshDavid05} we have found a new class of
supergravity cosmologies arising from time--dependent
compactifications of pure Einstein gravity, whose existence is
somewhat counter to the intuition of Ref.\ \cite{TW}. In
particular, we have exhibited exact solutions which circumvent the
no--go theorem \cite{nogo1}, while retaining {\it Ricci--flat}
internal spaces. This is possible through the introduction of one
or more geometric twists in the compact space. The observation
that acceleration is possible even for Ricci-flat cosmological
compactifications is not new, which only required the introduction
of external fluxes or
form-fields~\cite{OhtaPRL03,Emparan03a,IPN03d}. Our approach in
\cite{IshDavid05} was different in the regard that acceleration is
possible even if all available eternal fluxes are turned-off: in a
sense, the effect of form-fields is replaced by a non-trivial
``twist" in the internal geometry.
Here we considerably extend our knowledge of these solutions by
presenting new examples, and discuss their properties in a more
general setting.

The rest of the paper is organized as follow.
In Sec.~II, we review the product space compactification and further
discuss solutions to the vacuum Einstein equations where the
internal product space involves one or more geometric twists,
or alternatively some non-trivial curved internal spaces.
We examine these cosmological solutions broadly into three categories:
(i) a geometric twist alone; (ii) a non-trivial curvature alone, and
(iii) a combination of both. In the last case, (iii),
we will adopt a different gauge to solve the equations fully.

In Sec.~III, we discuss the cosmology on product spaces by using
the effective four-dimensional scalars which parameterize the
radii of the internal spaces and show that when the internal space
includes one or more twists along with non-flat extra dimensions,
then a scalar potential arising from a cosmological
compactification possesses a local (de Sitter) minimum, whose
existence generically predicts a period of accelerated expansion
in the four-dimensional Einstein conformal frame. In Sec. IV, we
discuss the properties of the effective potential in a more
general setting, which relate our solutions to the solutions given
in literature in the context of canonical 4d gravity coupled with
multiple scalars in an exponential form. In Sec. V, we will
present more new examples where the physical 3-space expands
faster than the internal space.

\section{Product space compactification: Basic Equations}

The interest in time-dependent supergravity (or
S-branes~\cite{CMChen:2002}) solutions arises mainly
from the relation of this system to string/M--theory and from a
`no-go' theorem~\cite{nogo1} which applies if one does not
consider cosmological solution. In fact,
the physical three space dimensions as ordinarily perceived are
intrinsically time dependent, and if the fundamental description
of nature would involve 6 or 7 (hidden) extra dimensions of space,
as is required for string/M theory to work, then it is reasonable to
assume that the internal space is also time dependent.
We therefore consider model cosmologies described by a general metric
{\it ansatz} in the following form:
\begin{eqnarray}\label{11dmetricGen}
&&\ds_{4+n}^2 = e^{-2\Phi}\ds_4^2 +\sum_i r_i^2\,
e^{2\phi_i}\dd\Sigma^2_{m_i,\epsilon_i},
\end{eqnarray} where
the parameters $r_i$ define appropriate curvature radii, and
$\ds_4^2$ is the metric of the physical large dimensions in
the form
\begin{equation}\label{FRWmetric}
\ds_4^2 = - {a(u)}^{2\delta} {\dd u}^2 + {a(u)}^2\dd\Omega^2_{k,3},
\end{equation}
so that $\phi=\phi(u)$. Here $\de$ is a constant, the choice of
which fixes the nature of the time coordinate, $u$, and
$\dd\Sigma^2_{m_i,\epsilon_i}$ are the metrics associated with
$m_i$-dimensional Einstein spaces; the values of
$\epsilon_i=0,\,+1,\,-1$ correspond to the flat, spherical, or
hyperbolic spaces, respectively.

For the metric~(\ref{11dmetricGen}), the Einstein-Hilbert
action is
\begin{eqnarray}
I &=& \frac{1}{16\pi G_{4+n}} \int d^{4+n} x \sqrt{-g} R \nonumber \\
&=& \frac{1}{16\pi G_{4+n}} \int d^n x \sqrt{g_n} \int d^4 x
\sqrt{-g_4} \e^{-4\Phi} \e^{\sum_i m_i\phi_i}
\left[\frac{R_4}{\e^{-2\Phi}}+\sum_i\frac{R_{m_i}}{\e^{2\phi_i}}+
\cdots \right].
\end{eqnarray}
where $\sum_i m_i\equiv n$. The dots denote, apart from the
obvious kinetic terms for the volume moduli $\phi_i$, the other
possible contributions to the 4d effective potential in the
presence of fluxes and/or geometric twists in the extra
dimensions. When the size of the internal space changes, as in the
present case, since $\phi_i\equiv \phi_i(u(t))$, the effective
Newton constant becomes time dependent, in general. This is
however not preferable for a model of our universe. To remedy this
problem, one must choose a Einstein conformal frame in four
dimensions by setting~\cite{CHNW03b}
\begin{equation}\label{rel-Phi-phi}
\Phi=\frac{1}{2}\sum_i m_i \phi_i.
\end{equation}
The Newton's constant in four dimensions is then
time--independent. With~(\ref{rel-Phi-phi}), the
$(4+n)$ dimensional action simplifies into the sum of the 4d
Einstein-Hilbert action plus an action for the scalar fields
$\phi_i$, which determine the size of extra dimensions.

In the product space compactification case, as studied in detail
in~\cite{CHNW03b,CHNOW03b}, we have $m_i\geq 2$. However, in this
paper we focus primarily in the case where some of the internal
(product) spaces involve one or more geometric twists, so that
$m_i=1$ for some $i$'s. For a mathematical simplicity we work in
the gauge $\delta=3$, in (\ref{FRWmetric}). The $tt$- and
$xx$-components of the field equations are
\begin{eqnarray}
&& 2K-\Phi^{\prime\prime}
+3\left(\frac{a^{\prime\prime}}{a}-3\frac{{a^\prime}^2}{a^2}\right)
=0 \label{tt-compo}\\
&&\Phi^{\prime\prime} +\left(\frac{a^\prime}{a}\right)^2
-\frac{a^{\prime\prime}}{a} - {2k}\,{a^4}= 0,\label{xx-compo}
\end{eqnarray}
where prime denotes a derivative with respect to $u$, $k$ is the
spatial curvature and
\begin{equation}\label{GeneralK}
K=\sum_{i=1} \frac{m_i(m_i+2)}{4} {\phi_i^\prime}^2
+\sum_{i>j=1} \frac{m_i m_j}{2}{\phi_i}^\prime
{\phi_j}^\prime\,.
\end{equation}
In the case of compactification on symmetric spaces taken in
direct products, the (scalar) field equations associated with
the extra dimensions are
\begin{equation}
\phi_i^{\prime\prime}+ (m_i-1)\sigma_i
e^{-(m_i+2)\phi_i-\sum_{i\neq j} m_j\phi_j} a^6=0,
\end{equation}
where $\sigma_i=\epsilon_i/r_i^2$. However, with the introduction of a
geometric twist along the internal space, the field equations for
$\phi_i$ will be modified.


To be more precise, let us consider a $(4+n)$ dimensional metric {\it
ansatz}, with $n=p+2q+1$,
\begin{equation}\label{11dmetricSpe}
\ds_{4+n}^2 = e^{-2\Phi}\ds_4^2 +
r_1^2\, e^{2\phi_1}\ds^2(\MM p_{\epsilon_1})+\ds_{2q+1}^2,
\end{equation}
where $\MM p$ is a $p$--dimensional space of constant
curvature of sign $\epsilon_1=0,\pm1$. The remaining $2q+1$
dimensions form a twisted product space
$$\MM{2q+1}=\underbrace{\MM 2\times \MM 2 \times \cdots
\times \MM 2}_{q~\text{times}} \ltimes S^1,$$ whose metric may be
given in the form
\begin{equation}\label{Twisted1}
\ds_{2q+1}^2= r_2^2 \e^{2\phi_2}\ds^2
\left({\cal M}_{\epsilon\Z2}^{2q}\right) +
r_3^2 e^{2\phi_3}\left(dz + f
\varpi_{\epsilon_2} \right)^2,
\end{equation}
with $f$ being the twist parameter.
More specifically,
\begin{equation}
\vbox{\halign{&$\displaystyle#$\hfil\cr
\ds^2(\MM{2q}\Z{+1})&=\sum_{i=1}^q(\dd
x_i^2+\sin^2x_i\,\dd y_i^2),\quad \varpi\Z{+1}=\cos x_i\,dy_i\hfil, \cr
\ds^2(\MM{2q}\Z0)\hfil&=\sum_{i=1}^q(\dd x_i^2+\dd y_i^2), \quad
\varpi\Z0=\frac{1}{2}(x_i dy_i- y_i
dx_i), \cr\ds^2(\MM{2q}\Z{-1})&=\sum_{i=1}^q(\dd x_i^2+\sinh^2x_i\,\dd
y_i^2),
\quad
\varpi\Z{-1}=\cosh x_i\,dy_i\cr}}.
\end{equation}
In the gauge $\delta=3$, the field equations for the scalars, with
arbitrary curvature $\epsilon_i$, may be given by
\begin{eqnarray}\label{scalar-equations}
&&\phi_1^{\prime\prime}+(p-1)\sigma_1 {\rm
e}^{-(p+2)\phi_1-2q\phi_2-\phi_3} a^6=0,\nonumber \\
&&\phi_2^{\prime\prime}+ \sigma_2 {\rm
e}^{-p\phi_1-2(q+1)\phi_2-\phi_3} a^6 -2F^2 {\rm
e}^{-p\phi_1-2(q+2)\phi_2+\phi_3} a^6 =0,\nonumber \\
&&\phi_3^{\prime\prime}+{2q} F^2 {\rm
e}^{-p\phi_1-2(q+2)\phi_2+\phi_3} a^6=0,
\end{eqnarray}
where $\sigma_1=\epsilon_1/r_1^2$, $\sigma_2=\epsilon_2/r_2^2$ and
$F\equiv (f/2) (r\Z2/r\Z3^2)$.

\subsection{Compactification on symmetric spaces}

Let us consider the case where $\sigma_i=0$ and $F^2=0$ (i.e.
without a geometric twist). The scalar wave equations then reduce
to $\phi_i^{\prime\prime}=0$. In this case one can easily solve
the Eqs.~(\ref{tt-compo}),(\ref{xx-compo}), with $k=0$. First
consider a special case where $\phi_1=$const. The corresponding
solution is
\begin{equation}
a=a_0 \e^{c\,u}, \quad \phi_1=\mbox{const}, \quad \phi_2=
\frac{2\sqrt{3}\,\ln{a}}{\sqrt{3\nu^2+{4}q(q+1+\nu)}}
=\phi_0+\frac{\phi_3}{\nu},
\end{equation}
where $a\Z0$, $\phi_0$, $c$ and $\nu$ are constants.
(Of course, the result does not depend on
$p$ in this case).

For $q\gg 1$, the decompactification of some of the extra
dimensions (and compactification of the other) can be a slow
process. This can be seen also from the solution below, where all
harmonic functions $\phi_i$ (or volume moduli) are time-dependent,
\begin{equation}
a=a\Z0 \e^{c u}, \quad \nonumber \phi_3=\mu \phi_2, \quad
\phi_1=\nu \phi_2, \quad \phi_2=\pm \,\eta \ln {a},
\end{equation}
up to a (different) shift in $\phi_i$, where
\begin{equation}
\eta\equiv \frac{2\sqrt{3}}{\sqrt{3\nu^2+4q(q+1+\nu)+
p(p+2)\mu^2+p\mu(\nu+2q)}}.
\end{equation}
The constants of integration $\mu,\nu, \eta$ may be chosen such
that each internal space shrinks with time.

Let us briefly discuss some common features of the above solution
by specializing it to a particular model with $p=4$ and $q=1$, so
that
$$ \Sigma_7=\TI^4\times \TI^2\times S^1.$$ As a particular case,
let us take $\phi_3=-4(\phi_1+\phi_2)=\phi_0=\text{const}$. The
corresponding solution is then given by
\begin{equation}
a(u)=a_0\,e^{c u}, \quad \phi_1=- \frac{\sqrt{3}}{2}
\ln(a(u)) =-(\phi_2+\phi_0/4).
\end{equation}
In this example, the $2$-space $\TI^2$ will grow with time, while
the $4$-space $\TI^4$ will shrink (or vice versa). Unfortunately,
since the acceleration parameter is always negative,
$\ddot{a}/a=-2c^2<0$ (an overdot denotes a derivative with respect
to cosmic time $t$, which is defined by $t= \int a(u)^3 \dd u$),
there is no acceleration in 4d Einstein conformal-frame. This all
imply that there is no de Sitter (accelerating) solution in
compactifications of a pure supergravity on maximally symmetric
spaces of zero curvature~\cite{TW}. However, inclusion of a
geometric twist along some of the internal (product) spaces, that
is crucial to our construction below, circumvents those arguments.

\subsection{The solution with a non-zero twist}

Consider the case where $\sigma_i=0$ (or $\epsilon_i=0$) but
$f>0$. The equations~(\ref{scalar-equations}) then reduce to
$\phi_1^{\prime\prime}=0$, $\phi_3^{\prime\prime} =-q
\phi_2^{\prime\prime} $. We can then solve the
equations~(\ref{tt-compo}) and (\ref{xx-compo}) simultaneously.
The explicit exact solution, with arbitrary values of $p$ and $q$,
is given by
\begin{eqnarray}\label{general-q-soln}
&&a=a\Z0 \left(\cosh\chi(u-u\Z0)\right)^{b\Z0} {\rm e}^{c\Z0 u},
\quad \phi_1= \frac{2}{p}\ln\left(\frac{f r\Z3 a\Z0^3}{\chi
r\Z2^2}\right) -c\Z1 u,\nonumber \\
&&\phi_2=\frac{1}{2}\ln\cosh\chi(u-u\Z0)+d\Z0 u
=-\frac{\phi_3}{q}, \label{q-twists}
\end{eqnarray}
where $b\Z0\equiv q/4$,
\begin{eqnarray}
&&c\Z0\equiv \frac{q(3qc+p c\Z1+4c)}{8}, \quad d\Z0\equiv
\frac{3qc+p c\Z1}{4}, \quad \chi\equiv\frac{1}{\sqrt{q}}\sqrt{8
c\Z0 d\Z0 -p(p+2)c\Z1^2-pq c c\Z1}\label{parameters}
\end{eqnarray}
and $a_0$ and $c$ are integration constants.
Note that not all constants of integration are shown: different
(gauge) shifts may be taken in $\phi_i$; some of which can be absorbed
into the $r_i$. The value of $u\Z0$ is merely a gauge choice
and so we set $u\Z0=0$ henceforth.
The solution~(\ref{general-q-soln}) would be
available even if $p=0$, in which case as there is no space ${\cal
M}^p$, the expression for $\phi_1$ drops off.

The 4d cosmic time $t$ is defined by
$\dd t=\pm a(u)^3\dd u$. The acceleration parameter is given by
\begin{equation}
{a^5\ddot a}= b\Z0 \chi^2-2 c\Z0^2 - 4 b\Z0 c\Z0\chi \tanh\chi u
- b\Z0 \chi^2 (1+2b\Z0)\tanh^2\chi u,
\end{equation}
where an overdot denotes differentiation w.r.t.\ cosmic time, $t$. It
follows that solutions will exhibit a
period of transient acceleration provided that
\begin{equation}
2 c\Z0^2-b\Z0 (1+2b\Z0)\chi^2 > 0.
\end{equation}
Acceleration occurs on the internal $u_-<u< u_+$, where
\begin{equation}
\tanh\chi u_{\pm}=\frac{-2c\Z0\pm
\sqrt{\chi^2(1+2b\Z0)-2c\Z0^2/b\Z0}}{\chi (1+2b\Z0)}.
\end{equation}
The number of e--folds during the (transient) period of
acceleration, $N_e\equiv \ln\,a(u_+)-\ln\,a(u_-)$ is given by
\begin{widetext}
\begin{equation}
N_e=\frac{b\Z0}{2}\ln \left[
\left(\frac{1+\tanh\chi u_+}{1+\tanh\chi u_-}\right)^{\kappa-1}
\left(\frac{1-\tanh\chi u_-}{1-\tanh\chi u_+}\right)^{\kappa+1}\right].
\end{equation}
where $\kappa\equiv c\Z0/(\chi b\Z0)$.
To the parameters
given in~(\ref{parameters}),
one has $\kappa\equiv ((3q+4)c+p\,c\Z1)/{2\chi}$.
The number of e--folds $N_e$
reaches a maximum when $c\Z1=0$, or equivalently $\ka=\sqrt{3q+4\over3q}$,
independently of $p$. This reflects
the fact that the $c\Z1=0$ solution is formally equivalent to the $p=0$
solution with no $\MM p$ torus and just the twisted $(2q+1)$--dimensional
space as an internal space. One may show that the maximum number of
e--folds is therefore given analytically by
\beq
(N_e)\ns{max}=\frac{q}8\ln\left\{\left((q+2)\sqrt{3}+2-\sqrt{q(3q+4)}
\over(q+2)\sqrt{3}-2-\sqrt{q(3q+4)}\right)^{\sqrt{3q+4\over3q}-1}
\left((q+2)\sqrt{3}+2+\sqrt{q(3q+4)}
\over(q+2)\sqrt{3}-2+\sqrt{q(3q+4)}\right)^{\sqrt{3q+4\over3q}+1}\right\}.
\eeq
\end{widetext}
The number of e-folds increases only marginally
when $q$ is increased.

Let us first briefly discuss some common features of the $f>0$
solution by setting $q=2$, so that
\begin{equation}
b\Z0=1/2, \quad c\Z0=(10c+pc\Z1)/{4}=d\Z0+c, \quad \chi = \sqrt{15
c^2+3 p c c\Z1- p(4+p) c_1^2/4}.
\end{equation}
The solution will then exhibit a period of transient acceleration
provided that $\al\Z{2-}c<c\Z1<\al\Z{2+}c$, where
\begin{equation}
\al\Z{2\pm}=\frac{2p\pm 4\sqrt{2p(2p+5)}}{p(3p+8)}.
\end{equation}
This implies, e.g. when $p=2$,
$-\frac{5}{7} c< c_1 <c $.
Acceleration occurs on the internal $|u_-|<u<|u_+|$ where
\begin{equation}
\tanh(\chi u_\pm )=\frac{-\kappa\pm \sqrt{2-\kappa^2}}{2},
\end{equation}
and $\kappa\equiv (10c+pc\Z1)/2\chi$. The volume moduli
associated with ($\TI^2\times \TI^2)\ltimes S^1$ are stabilized at
late times when
$$
c_1= \pm \frac{6c}{\sqrt{3p(p+2)}}.$$ For $p=2$, this yields $c_1=
\pm \sqrt{\frac{3}{2}} c $, which lies outside the range that is
required for acceleration. In the above example, the extra
dimensions cannot be stabilized completely, if there is to exist a
period of acceleration in the four-dimensional Einstein conformal
frame.

\subsection{Solutions with flat and hyperbolic extra dimensions}

In this subsection we focus primarily on solutions that are
obtained from compactifications on a product of hyperbolic and
flat spaces. In fact, in the $F^2=0$ case, to have a positive
potential we need to compactify a higher dimensional gravitational
theory on spaces that include, at least, one negatively curved
factor~\cite{TW,CHNOW03b}. Here we can make the problem simpler by
considering that (i) the $p$ dimensional space $\MM p$ is
negatively curved while the $2q$-space $\MM{2q} =\MM{2}\times
\cdots \times \MM 2$ is Ricci flat ($\MM 2= \TI^2$) (case I) or
(ii) $\MM p$ is Ricci flat and $\MM{2q}$ is negatively curved
($\MM 2 = \HI^2$) (case II).

In the first case, the solution is
given by
\begin{eqnarray}\label{FirstSol}
&&a(u)=\frac{\e^{c\Z0 u}}{\left(\sinh\chi
{|u|}\right)^{b\Z0}},\quad
\phi_1= c\Z1 {u}-\frac{1}{(p-1)}\ln\sinh\chi{|u|}, \nonumber \\
&&\phi_2=c\Z2 {u}, \quad \phi_3=c\Z3 {u}+
2\ln\left(\frac{p-1}{\chi r\Z1}\right),
\end{eqnarray}
where $b\Z0\equiv  p/[2(p-1)]$ and
\begin{eqnarray}\label{first-case}
&& c\Z0\equiv \frac{p((p+2)c+2q c\Z2+c\Z3)}{4(p-1)},\quad
c\Z1\equiv \frac{3 c p +2 q c\Z2 +c\Z3}{2(p-1)},
\nonumber \\
&&\chi^2=\frac{1}{4p}\left[4q c\Z2 (3p-2)(c\Z3+c\Z2 q) +8q(p-1)
c\Z2^2 +(7p-6)c\Z3^2 + 3p^2\left((p+2)c^2+4q c c\Z2+2c
c\Z3\right)\right],
\end{eqnarray}
with $c$ being an arbitrary constant.

In the second case, the explicit solution is
\begin{eqnarray}\label{SecondSol}
&&a(u)=\frac{{\rm e}^{c\Z0 u}}
{\left(\sinh\chi{|u|}\right)^{b\Z0}}, \quad \phi_1=c\Z1 {u}, \quad
\phi_2= {c\Z2} {u}
-\frac{1}{2q-1}\ln\sinh\chi{|u|}, \nonumber \\
&& \phi_3=c\Z3 {u}+ 2\ln\left(\frac{\sqrt{2q-1}}{\chi
r\Z2}\right),
\end{eqnarray}
where $b\Z0\equiv q/(2q-1)$ and
\begin{eqnarray}\label{second-case}
&& c\Z0\equiv
\frac{q(pc\Z1+c\Z3+2(q+1)c)}{2(2q-1)}, \quad
c\Z2\equiv \frac{pc\Z1 +c\Z3+6qc}{2(2q-1)},\nonumber \\
&&\chi^2=\frac{1}{4q}\left(p\left[q(3p+4)-p-2\right]c\Z1^2 +12 q^2
(q+1) c^2 + 12 q^2(p c\Z1+ c\Z3)c + 2(3q-1)pc\Z1 c\Z3 + (7q-3)
c\Z3^2\right).
\end{eqnarray}
The solutions given in Ref.~\cite{CHNOW03b} may be realized as
some limiting cases of the solutions presented above, and fall in
the category of accelerating cosmologies discussed, e.g.
in~\cite{TW,CMChen:2002}. In the case that $\HI^{p}$ is replaced
by $S^{p}$ (or $\HI^{2q}$ by $S^{2q}$), the function $\sinh$ would
be replaced by $\cosh$. In this case it is suggestive to study the
possible non-perturbative (instanton) effects, which may help to
decrease the slope of the runway potential. We should however note
that, in the case $\MM p\to \HI^{p}/\Gamma$, there are no moduli
other than the volume modulus, because only modulus of a compact
hyperbolic Einstein space of dimensions $p\geq 3$ is its volume;
for a discussion, see, e.g.~\cite{Kaloper00,IPN03d} and references
therein.

Let us write the scale factor in a canonical form:
\begin{equation}
a(u)= \frac{\e^{c\Z0 u}}
{\left(\sinh(\chi{|u|})\right)^{b\Z0}}.
\end{equation}
The expansion parameter is
\begin{equation} \label{HubblePara1}
\dot{a}
=\pm \frac{1}{a^2}
\left(c\Z0 -\chi b\Z0 \coth\chi{u}\right)\equiv \pm
\frac{a^\prime}{a^3},
\end{equation}
depending upon the choice of sign in ${t} \equiv \pm \int
a^3(u) {\dd u}$,
while the acceleration parameter is given by
\begin{equation}\label{AccelPara1}
\ddot a= \frac{1}{a^5}\bigg(-b\Z0 \chi^2-2 c\Z0^2 + 4 b\Z0
c\Z0\chi \coth\chi u + b\Z0 \chi^2 (1-2b\Z0)\coth^2\chi u\bigg).
\end{equation}
The four-dimensional universe is expanding if $\dot{a}>0$. The
universe is undergoing accelerated expansion if, in addition,
$\ddot{a}>0$. It follows from~(\ref{HubblePara1}) that $\dot{a}$
cannot change sign when $u<0$ (upper sign) or $u>0$ (lower sign),
since $\chi b\Z0>c\Z0$. And the solution will exhibit a period of
transient acceleration in 4d Einstein frame provided that $2
c\Z0^2 +b\Z0 (1-2 b\Z0) \chi^2 >0$. Acceleration occurs on the
interval $u_+<u<u_-$, where
\begin{equation}
\coth(\chi {u_\pm} )=\frac{2 c\Z0 \pm
\sqrt{\chi^2(1-2b\Z0)+2c\Z0^2/b\Z0}}{(2b\Z0-1)\chi}.
\end{equation}

\begin{figure}[htb]
\vbox{\vskip-10.5pt\centerline{\scalebox{0.75}{\includegraphics{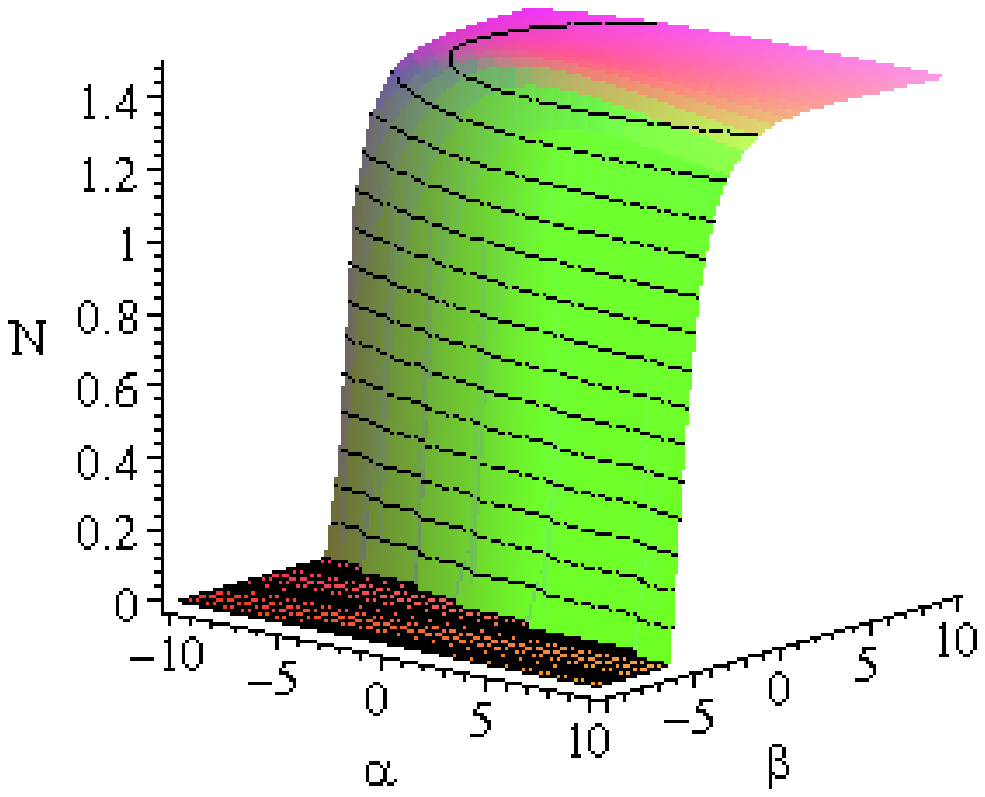}}}
\caption{\label{fig1}%
{\sl The number of e--folds during acceleration epoch (case I)
as a function of the parameter $\alpha\equiv c\Z2/c$, $\beta\equiv
c\Z3/c$, for $p=17$, $q=2$, ($d=26$). $N_e$ is maximum around
$(\alpha,\beta)=0$. }}}
\end{figure}

\begin{figure}[htb]
\vbox{\vskip-10.5pt\centerline{\scalebox{0.75}{\includegraphics{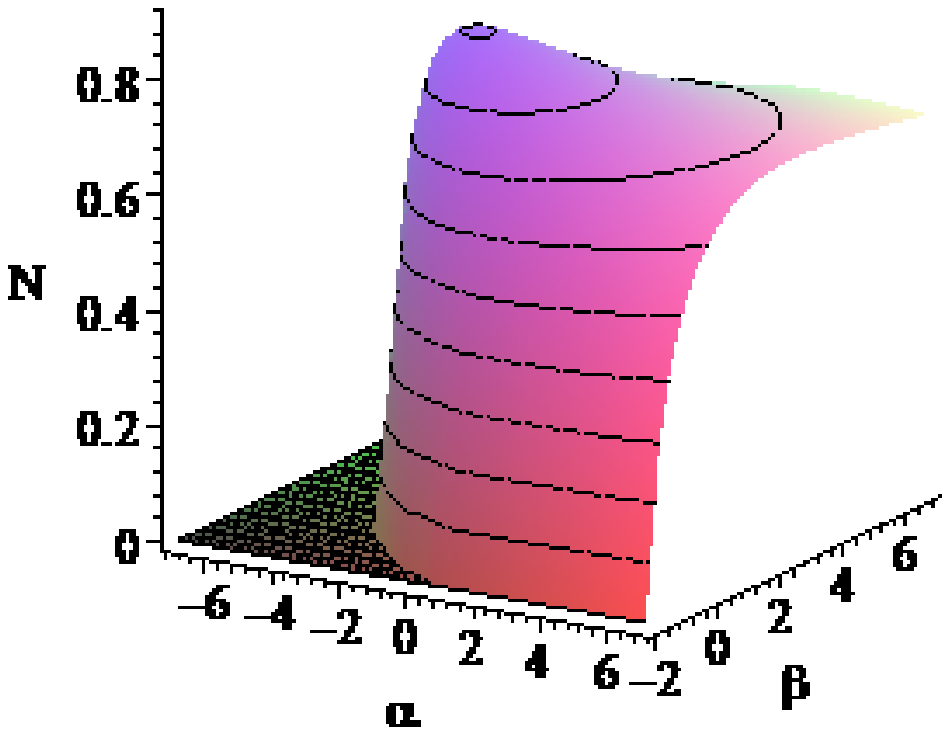}}}
\caption{\label{fig2}%
{\sl The number of e--folds during acceleration epoch (case II)
as a function of the parameter $\alpha\equiv c\Z1/c$, $\beta\equiv
c\Z3/c$, for $p=2$, $q=2$, ($d=11$). $N_e$ is maximum around
$(\alpha,\beta)=0$.}}}
\end{figure}

The number of e-folds is given analytically by
\begin{equation}
N_e=\ln \left\{
\left(\frac{\coth\chi{u_+}+1}
{\coth\chi{u_-}+1}\right)^{\kappa_+}
\left(\frac{\coth\chi{u_-}-1}
{\coth\chi{u_+}-1}\right)^{\kappa_-}\right\},
\end{equation}
where
\begin{equation}
\kappa_\pm =\frac{1}{2} \left(\frac{c\Z0}{\chi}\pm b\Z0\right).
\end{equation}
Intuitively, the (transient) acceleration occurs far from the
cosmological singularity at $u=0$.

To the parameters given in~(\ref{first-case})
or~(\ref{second-case}), the number of e-folds is of order unity,
see Figs.~\ref{fig1} and \ref{fig2}. However, this is not a
disaster, since the universe would have been expanding prior to
the period of transient acceleration, with the number of e--folds
\begin{equation}
\overline{N}_e
=\int \frac{\dot{a}}{a} \,
{\dd t}= \int_{\tilde{u}}^{u_-} \frac{a^\prime}{a}\dd u.
\end{equation}
If $\Delta u =|u_--\tilde{u}|\lesssim {10}^{-36}$, then it is not
impossible that the total number of e-folds $\overline{N}_e+N_e >
60$, as required to explain the flatness problem. In our model, it
is possible that the scale factor of our universe remained much
larger than the size of internal dimensions at the beginning of
the accelerating epoch.

\subsection{The field equations in $\delta=0$ gauge}

Let us consider the case where, at least, one of the $\sigma_i$ is
non-zero. We also demand that $f>0$. In this case we find it
convenient to choose the gauge $\delta=0$ in the metric {\it
ansatz}~(\ref{FRWmetric}), so that $u$ becomes the (proper) cosmic
time, $t$, and $\phi_i \equiv \phi_i(t)$. Then upon dimensional
reduction from $(4+n)$ dimensions to four-dimensions, we get
\begin{equation}
I=\frac{1}{8\pi G_{4}}\int {d^4 x} \sqrt{-g_4}\left( \frac{R_4}{2}
+K -V \right),
\end{equation}
where $G_4=G_{4+n}/\text{Vol}({\cal M}^n)$. The kinetic term $K$
is given by~(\ref{GeneralK}), after replacing the time derivative
$'=\partial/\partial{u}$ by $\dot{}=\partial/\partial{t}$, along
with the substitutions: $m_1= p$, $m_2= 2q$ and $m_3= 1$. While,
the potential term is given by
\begin{equation}\label{FullPotential}
V = \Lambda_1 e^{-(p+2)\phi_1-2q\phi_2-\phi_3} + \Lambda_2
e^{-p\phi_1-2(q+1)\phi_2-\phi_3} + q{F}^2
e^{-p\phi_1-2(q+2)\phi_2+\phi_3} \equiv V_1 +V_2 + V_F,
\end{equation}
where $\Lambda_1=-\epsilon_1 \frac{p(p-1)}{2 r\Z1^2}$,
$\Lambda_2=-\epsilon_2\frac{q}{r\Z2^2}$ and $F=(f/2)
(r\Z3/r\Z2^2)$. The analogue Einstein equations are
\begin{eqnarray}
&&\ddot{\Phi}+3\frac{\dot{a}}{a}\dot{\Phi}-2
K-3\frac{\ddot{a}}{a}=0,
\label{(4+n)Eq1} \\
&&\ddot{\Phi}+3\frac{\dot{a}}{a}\dot{\Phi} -2\left(\frac{\dot{a}}{a}\right)^2
-\frac{\ddot{a}}{a} - \frac{2k}{a^2} = 0, \label{(4+n)Eq2} \\
&&\ddot{\phi_1}+3\frac{\dot{a}}{a}\dot{\phi}_1-\frac{2}{p} V_1 = 0, \\
&&\ddot{\phi_2}+3\frac{\dot{a}}{a}\dot{\phi}_2-\frac{1}{q}V_2-\frac{2}{q} V_F=0, \\
&&\ddot{\phi_3}+3\frac{\dot{a}}{a}\dot{\phi}_3+ 2 V_F = 0,
\end{eqnarray}
where $\Phi=p\phi_1/2+q\phi_2+\phi_3/2$. In terms of alternative
canonically normalized scalars, which may be defined by
\begin{eqnarray}
&&\varphi_1=\sqrt{\frac{q}{2(q+1)}}\left(p\phi_1+2(q+1)\phi_2
+\phi_3\right), \nonumber \\
&& \varphi_2=\sqrt{\frac{p(p+2q+2)}{2(q+1)}}\left(\phi_1
+\frac{1}{p+2q+2}\phi_3\right),\nonumber \\
&& \varphi_3=-\sqrt{\frac{p+2q+3}{p+2q+2}}\phi_3,
\end{eqnarray}
the field equations take the following form
\begin{equation}
\ddot{\varphi_i}+3H\dot{\varphi_i}+\frac{dV}{d\varphi_i}=0,
\label{MainWave1} \end{equation}
\begin{equation}
 \dot{H}+K-k a^{-2}=0,\label{MainKH1}
\end{equation}
along with the Friedmann (constraint) equation
\begin{equation}
H^2=\frac{1}{3} \left(K+V\right)-k a^{-2},\label{MainCon1}
\end{equation}
where $K=\frac{1}{2}\sum_i^3 \dot{\varphi_i}^2$. The resulting
scalar potential is
\begin{equation}\label{MainPoten1}
V=\Lambda_1 \e^{-\beta_1\varphi_1-\beta_2\varphi_2}
+\Lambda_2 \e^{-2\varphi_1/\beta_1}
+qF^2 \e^{-\beta_3\varphi_1+\beta_4\varphi_2-
\beta_5\varphi_3},
\end{equation}
where
\begin{eqnarray}
&&\beta_1 \equiv \sqrt{\frac{2q}{q+1}}, \quad \beta_2\equiv
\sqrt{\frac{2(p+2q+2)}{p(q+1)}},\quad \beta_3 \equiv
\frac{q+2}{q}\beta_1,\nonumber \\
&&\beta_4 \equiv \frac{2}{q+1}\frac{1}{\beta_2},\quad \beta_5
\equiv 2\sqrt{\frac{p+2q+3}{p+2q+2}}.
\end{eqnarray}
Clearly the potential has a positive definite minimum with respect
to a subset of the $\varphi_i$ directions. By comparing the $f=0$
and $f>0$ potentials, we see that a geometric twist has a more
significant influence on shape of the potential. In general, a
scalar potential as above, with cross coupling exponents, implies
the existence of a local de Sitter region. Thus one may expect
that accelerating cosmologies with a larger number of e-folds is
possible in this case.

It is rather non-trivial to write the general solution of
equations (\ref{MainWave1})-(\ref{MainCon1}) with the potential
(\ref{MainPoten1}), in terms of proper (cosmic) time $t$, but any
such solutions should be the same as the one from $4+n$
dimensional field equations. Accepting that the effective
potential is a useful tool in this investigation, in later
Sections we present various explicit solutions that correspond to
the above set of equations, but in terms of a new time coordinate,
$\tau$.

\subsection{M-theory phantom cosmology}

Let us first briefly discuss a special case where the scalar
potential $V(\varphi)=0$. With $\Lambda_i=0=F^2$, the solution for
the scalars is
\begin{equation}
\dot{\phi_i}(t)=\sqrt{3} \frac{c_i\,a_0^2}{a(t)^3},
\end{equation}
where a dot denotes differentiation with respect to cosmic time
$t$. The factor of $\sqrt{3}$ is introduced just for a
convenience. The corresponding kinetic term is
\begin{equation}
K=3C\frac{{a_0}^4}{a(t)^6},\end{equation} where
\begin{equation}
C\equiv 6c_1^2+2c_2^2+\frac{3}{4}c_3^2+4 c_1 c_2+c_2 c_3+2 c_1
c_3.
\end{equation}
The effective potential, say $\tilde{V}$, can have a non-zero
contribution if the intrinsic curvature of the physical 3-space is
non-zero, $k\neq 0$, see e.g.~\cite{CHNW03b}. In the $k=-1$ case,
the solution for the scale factor is implicitly given by
\begin{equation}
\int_{}^{a(t)} \frac{x^2}{\sqrt{x^4+C a_0^4}} dx = t+t_1.
\end{equation}
where $t_1$ is an arbitrary constant. The $C=0$ solution is
special, as it implies $a(t)=t+t_2$
and $\ddot{a}=0$. This solution critically differentiates between
eternally accelerating expansion (if $C<0$) and decelerating expansion
($C>0$). In the case $K=V(\varphi)=0$, the
cosmological trajectories can be the null geodesics in
a Milne patch ($k=-1$) of 4d Minkowski spacetime,
following~\cite{Townsend:2004zp}

The choice where all $c_i$ are zero except one corresponds to
phantom cosmology, if the non-vanishing $c_i$ is imaginary. In
this case, since $C<0$ (or $K<0$), the corresponding solution
yields $\mbox{w} <-1$, where $\mbox{w}\equiv
(K-\tilde{V})/(K+\tilde{V})$, and also that $\mbox{w}\to -1$ as
$t\to \infty$ (for a discussion of phantom cosmology with a single
scalar in exponential form, see, e.g.
Refs.~\cite{Sami:2003xv,Nojiri0405}). In our model, however, the
choice $C<0$ is not physically motivated, because in this case
some of the extra dimensions behave as time-like, rather than
space-like.

\section{Multiple scalars and cosmology in four-dimensions}\label{4dMultiCos}

In this Section, we give a general discussion on solving scalar
field equations with multiple exponential potentials in a flat FRW
universe. First, we note that for the compactification of
classical supergravities on product spaces, with or without
certain `twists' in the geometry twist, one can always bring the
scalar potential into the form
\begin{equation}\label{mainpoten}
V(\varphi) = \sum_{i=1}^{s} \Lambda_i \,\e^{- \frac{1}{M_P} \sum_{i
\geq j \geq 1}^{i+j-1} \lambda_{i+j-1} \varphi_j},
\end{equation}
where $\varphi_i$ are canonically normalized 4d scalars.
The kinetic term is given by
$K=\frac{1}{2}\sum_{i}^{s}\dot{\varphi}_i^2$. The corresponding
equations of motion for the scalar fields are
\begin{equation}
\ddot{\varphi}_i+3H\dot{\varphi}_i
+\frac{dV}{ d\varphi_i}=0, \label{waveeq}
\end{equation}
while the Friedmann equation is
\begin{equation}\label{Fried1}
H^2= \frac{\rho_\varphi}{3M_P^2}-\frac{k}{a^2},
\end{equation}
where $a$ is the scale factor, $H\equiv \dot{a}/a$ is the Hubble
parameter, which represents the universal rate of expansion,
$\rho_\varphi\equiv K+V(\varphi)$ is the energy density of the
scalar fields, $k$ is the spatial curvature.

The values of (dilaton) coupling constant $\lambda_i$ are model or
compactification scheme dependent, e.g. in the case of two scalars
arising from hyperbolic compactification of (4+m) dimensional
supergravity~\cite{CHNW03b}, we find $\lambda_1 ={2}/{\lambda_2
}=\pm \sqrt{2+4/m_1}$ and $ \lambda_3 =
2\sqrt{(m_1+m_2+2)/[m_2(m_1+2)]}$. Further, $\Lambda_i=
-\epsilon_i \frac{m_i(m_i-1)}{2}\,
\left(\frac{M_P}{r_i}\right)^{2}$, so that $\epsilon_i=0, +1$ or $
-1$, respectively, for flat, spherical or hyperbolic space.

\subsection{The two scalar case}

Consider the potential with $2$ scalars $\varphi_1,\varphi_2$:
\begin{eqnarray}\label{potential1}
V(\varphi) &=& {\Lambda_1}\,
\e^{- {\lambda_1}\varphi_1}+{\Lambda_2}\,
\e^{-  {\lambda_2} \varphi_1- {\lambda_3}
\varphi_2} \end{eqnarray}
(in units $M_P^{-1}=1$). The exact solution with the single scalar
$\varphi_1$ has appeared in~\cite{Townsend03b} (Ref.~\cite{IPN03d}
provides further generalizations, and
Ref.~\cite{Chimento98a,jrusso04a} contain solutions in different
time coordinates). In the $\Lambda_2=0$ case, the field
$\varphi_2$ acts merely as a non-interacting massless scalar. This
case was studied in~\cite{Chimento98a}. Thus we focus here on
solutions with more than one scalars, where in general
$\Lambda_i\neq 0$. We present solutions with two and three
scalars, but the method would be equally applicable to higher
number of scalar fields.

In the two scalar case, the late time behavior of the scale factor
and the scalars is characterized by
\begin{equation}\label{scalefac1}
a(t) \propto \left\{\begin{array}{l}
(t/t_0)^{\gamma},
~~~~~~~~ \quad k=0 , \\ \\
\sqrt{|k|}\,t, ~~~~~~~~~ \quad k\neq 0,
\end{array} \right.
\end{equation}
\begin{equation}\label{twoscalars}
 \varphi_1=\frac{1}{\lambda_1}\left(2\ln{t}-\ln \arrowvert
{\varphi_1^{(0)}}\arrowvert \right), \quad
\varphi_2=\frac{1}{\lambda_1}\left(2\mu\ln {t}
+\frac{\lambda_2}{\lambda_3}\ln \arrowvert
{\varphi_1^{(0)}}\arrowvert -\frac{\lambda_1}{\lambda_3} \ln
\arrowvert {\varphi_2^{(0)}}\arrowvert \right),
\end{equation}
where
\begin{eqnarray}
&&\gamma \equiv \frac{2(1+\mu^2)}{\lambda_1 ^2}, \quad \mu\equiv
\frac{\lambda_1-\lambda_2}{\lambda_3}, \\
&&\varphi_1^{(0)}=\frac{2(3p-1)}{\Lambda_1\lambda_1^2}
\left(1-\frac{\lambda_2}{\lambda_3}\mu\right),\quad
\varphi_2^{(0)}= \frac{(3p-1)}{\Lambda_2}
\frac{2\mu}{\lambda_1\lambda_3}.
\end{eqnarray}
The late time attractor solutions, with $\Lambda_1=0$, have
appeared before in Ref.~\cite{Ed99a}. Since these asymptotic
solutions may not be taken to far, below we present a more general
class of solutions which are available before the attractors are
reached.

\subsection{Accelerating solutions}

With any number of scalars, there would be a period of
acceleration (where $\dot{a}>0,~\ddot{a}>0$) before approaching
the attractor provided that $\gamma>1/3$.
The universe inflates when one of the fields approaches its
minimum, where the energy density is dominated by the potential
energy of scalar fields. In particular, the solution around
$\varphi_1 =0$ and $\varphi_2 =\frac{c\Z0}{\lambda_3}$
($c\Z0$ is defined below) is given by
\begin{equation}\label{scalefactor1}
a(t) = \left\{\begin{array}{l}
a_{0}\,\e^{ H_0\, t},
~~~~~~~~~~~~~~~~~~~ \quad k=0 , \\ \\
(\sqrt{k}/H_0)\,\cosh {H_0\,t} , ~~~~~
\quad k>0 , \\ \\
(\sqrt{-k}/H_0)\,\sinh {H_0\,t}, ~~~ \quad k <0,
\end{array} \right.
\end{equation}
where the Hubble rate $H_0$ is characterized by the scale
\begin{equation}
H_0 = \sqrt{\frac{\gamma^2}{\varphi_1^{(0)}}}.
\end{equation}
In a flat universe the solution~(\ref{scalefactor1}) is held only
as an intermediate stage in the evolution unless $\gamma> 1$.
For $k\neq 0$, however, the period of acceleration depends on the
initial value of $\varphi_i$, other than the couplings
$\lambda_i$. The inflating $k=-1$ slice that approaches the
geodesic $a(t) \sim {t}$ asymptotically can be eternally
accelerating even if $\gamma<1$, as in the single scalar double
exponential case~\cite{Jarv04a,IPN03d}.

To find the exact solution, with $k=0$, one introduces a new
logarithmic time variable $\tau$, which is defined by
$$\alpha=\ln(a(t)), \quad \tau =\int^t \e^{-\lambda_1\varphi_1(\bar{t})/2}
\dd\bar{t}.$$ For $k=0$, the field
equations~(\ref{waveeq})-(\ref{Fried1}), with $i=1,2$, reduce to
\begin{eqnarray}
\alpha^\prime&=& \frac{1}{2\sqrt{3}}\sqrt{V_0} \left(\xi+\xi^{-1}\right),\\
\varphi_1^\prime &=&\frac{1}{\sqrt{2}}\sqrt{V_0}\sqrt{1+\mu^2}
\left(\xi-\xi^{-1}\right), \\
\varphi_2 &=& \mu \varphi_1 +\lambda\Z3^{-1} c\Z0,
\end{eqnarray}
where a prime denotes differentiation w. r. to $\tau$,
\begin{equation}
V_0\equiv \Lambda\Z1+\Lambda\Z2 \e^{-c\Z0}, \quad c\Z0\equiv
\ln \vert \frac{\Lambda\Z2}{\Lambda\Z1}\,
\frac{\lambda_3^2-\lambda_2(\lambda_1-\lambda_2)}
{\lambda_1(\lambda_1-\lambda_2)}\vert
\end{equation}
and the variable $\xi$ satisfies the differential equation
\begin{equation}
\xi^\prime =\sqrt{V\Z0}(4\gamma)^{-1/2}
\left((\sqrt{3\gamma}+1)-(\sqrt{3\gamma}-1)\xi^2\right).
\end{equation}
The $k=0$ hyperbola $3{\alpha^\prime}^2-(1+\mu^2)
{\varphi_1^\prime}^2/2=V_0$ separates the $k=+1$ and $k=-1$
trajectories. Though it is rather non-trivial to write the
solutions explicitly when $k=\pm 1$, one may extract the common physical
effects by studying the trajectories in a phase portrait,
following~\cite{Jarv04a,halliwell}.

Let us consider the $\alpha^\prime>0$ branch, which restricts the
solution to expanding cosmologies. For $\gamma>1/3$, the explicit
solution is
\begin{equation}\label{SolnForXi}
\xi=\sqrt{\frac{\sqrt{3\gamma}+1}{\sqrt{3\gamma}-1}} \tanh
\sqrt{V_0}\sqrt{\frac{3\gamma-1}{4\gamma}}\tau, \end{equation} up
to a shift of $\tau$ about $\tau=0$. Hence
\begin{eqnarray}
a&=&a_0 \left(\cosh\zeta\tau\right)^{\delta_-}
\left(\sinh\zeta\tau\right)^{\delta_+}, \label{GeneralSol1}
\\
\varphi_1 &=&\frac{1}{\beta}
\ln\left(\left(\cosh\zeta\tau\right)^{\delta_-}
\left(\sinh\zeta\tau\right)^{-\delta_+}\right)+\varphi_0,
\label{GeneralSol2}
\end{eqnarray}
where $\beta\equiv \sqrt{(1+\mu^2)/6}$,
\begin{equation}\label{delplusminus}
\delta_{\pm}=\sqrt{\frac{\gamma}{3}} \frac{1}{\sqrt{3\gamma}\pm
1}, \quad
\zeta= \sqrt{V_0}\sqrt{\frac{3\gamma-1}{4\gamma}}
\end{equation}
and $a_0$ and $\varphi_0$ are integration constants. To obtain the
solution for $\gamma<1/3$, one replaces $\tanh{X}$ by $\tan{X}$
in~(\ref{SolnForXi}). Since the scaling regime of exponential
potentials does not depend upon its compactification or mass
scales ($\Lambda_i$), $\varphi_0$ is actually a free parameter
that can, for simplicity, be set to $M_P$ or even to zero.

The acceleration parameter is given by
\begin{equation}\label{4dCosmoAcce}
\frac{\ddot{a}}{a}= {2\zeta^2}\,
\e^{-\sqrt{\frac{12}{\gamma}}\beta\varphi_1}
\left[4\delta_-\delta_+ -(\delta_-\tanh\zeta\tau)^2
-(\delta_+\coth\zeta\tau)^2\right],
\end{equation}
which decreases as $\tau$ increases. This actually implies that
the $\tau\to\infty$ limit corresponds to $t\to \infty$, to the
above solutions. It follows from~(\ref{4dCosmoAcce}) that, when
$\gamma<1$, the solution will exhibit a period of transient
acceleration. This occurs on the interval $\tau_-<\tau<\tau_+$,
where
\begin{equation}
(\tanh\zeta\tau_{\pm})^2=\frac{(2\pm
\sqrt{3})(\sqrt{3\gamma}-1)}{\sqrt{3\gamma}+1}.
\end{equation}
For future use, we also note that the
shift in $\varphi_1$ during the accelerated epoch is given by
\begin{equation}
\delta\varphi_1=\frac{1}{\beta}\left[\delta_-\ln\frac{\cosh\zeta\tau_+}{\cosh\zeta\tau_-}
-\delta_+\ln\frac{\sinh\zeta\tau_+}{\sinh\zeta\tau_-}\right],
\end{equation}
while $\delta\varphi_2=\mu \delta\varphi_1$. At late times,
$\tau\to \infty$ (or $t\to \infty$), we find
\begin{equation}
\frac{\ddot{a}}{a}\propto \frac{\gamma(\gamma-1)}{(3\gamma-1)^2}.
\end{equation}
An important difference between $\gamma>1$ and $\gamma=1$
solutions is that in the formal case the number of e-folds is
arbitrary, while in the latter it is fixed, $N_e=11.68$. Thus, the
$\gamma=1$ solution may be viewed as a transient only since it has
got a natural entry and exit from inflation; the period of
acceleration can be made arbitrary large but not the e-folds!

The number of e-folds $N_e$ during the accelerated epoch
is given by
\begin{equation}
N_e=\int_{t_i}^{t_f} H\dd t= \int_{\tau_-}^{\tau_+}
\frac{a^\prime}{a}\dd\tau = \int_{\tau_-}^{\tau_+} \zeta
\left(\delta_- \tanh\zeta\tau +\delta_+ \coth\zeta\tau\right),
\end{equation}
where $t_i$ and $t_f$ are the starting and ending times of the accelerated
expansion. The parameters $\delta_{\pm}$ and $\zeta$ are constants
which were defined previously in~(\ref{delplusminus}).

For $\gamma<1$, the number of e-folds during an acceleration epoch is
given analytically by
\begin{equation}
N_e=\frac{\gamma}{3\gamma-1}\ln
\frac{(3-\sqrt{3})(1+\sqrt{\gamma})}{(3+\sqrt{3})(1-\sqrt{\gamma})}
+\sqrt{\frac{\gamma}{3}}\frac{1}{\sqrt{3\gamma}-1}
\ln\frac{\sqrt{3}+1}{\sqrt{3}-1}.
\end{equation}
As some specific values, one has $N_e=2.62$ for $\gamma=0.99$;
$N_e=1.5$ for $\gamma=0.9$; $N_e=1.16$ for $\gamma=0.8$. It is
worth noting that the number of e-folds depends only upon the
effective value of the acceleration parameter $\gamma$, but not on
the number of scalar fields. 

In the $\gamma\leq 1$ case, both the
lower and the upper limits ($\tau_-$ and $\tau_+$) are fixed in
terms of $\zeta$ and $\gamma$, while, in the $\gamma>1$
case, only the lower limit of integration (the on-set time of
accelerated epoch) is fixed, which is given by
\begin{equation}
\tau_-=\frac{1}{\zeta} \tanh^{-1} \sqrt{\frac{(2-\sqrt{3})
(\sqrt{3\gamma}-1)}{\sqrt{3\gamma}+1}}.
\end{equation}

\subsection{The three scalar case}

It is straightforward to generalize the above results with higher
number of scalar fields. Let us consider the potential with
$3$ scalars $\varphi_1,\varphi_2, \varphi_3$, which is given by
\begin{equation}\label{potential2}
V = {\Lambda_1}\,
\e^{- \lambda_1 \varphi_1}+ {\Lambda_2} {\rm
e}^{- \lambda_2\varphi_1-\lambda_3 \,\varphi_2}+
{\Lambda_3}\,{\rm
e}^{-\lambda_4\varphi_1-\lambda_5\varphi_2-\lambda_6\varphi_3}.
\end{equation}
The solution to the equations (\ref{waveeq}) (\ref{Fried1}), with the
potential~(\ref{potential2}), is
given by~(\ref{GeneralSol1})-(\ref{GeneralSol2}) but now
\begin{equation}
\beta\equiv \frac{1}{\sqrt{6}} \sqrt{1+\mu^2+\nu^2},\quad
V\Z0\equiv \Lambda_1+\Lambda_2 \e^{-c\Z0}+\Lambda_3 {\rm
e}^{-d\Z0},
\end{equation}
where
\begin{eqnarray}
&&\nu\equiv \frac{\lambda_3(\lambda_1-\lambda_4)-
\lambda_5(\lambda_1-\lambda_2)}{\lambda_3\lambda_6}, \quad
\gamma=\frac{2}{\lambda_1^2}\left(1+\mu^2+\nu^2\right),
\nonumber \\
&&c\Z0\equiv
d\Z0-\ln\left|\frac{\Lambda_3}{\Lambda2}\,
\frac{\mu\lambda_6-\nu\lambda_5}{\nu\lambda_3}\right|,
\nonumber \\
&&d\Z0 \equiv \ln \left| \frac{\Lambda_3}{\Lambda_1} \frac{
\lambda_6(\lambda_3-\mu\lambda_2)+\nu(
\lambda_2\lambda_5-\lambda_3\lambda_4)}{\lambda_1\lambda_3}\right|,
\end{eqnarray}
and
\begin{equation}
\varphi_2=\mu\varphi_1+\lambda_3^{-1} c\Z0, \quad
\varphi_3=\nu\varphi_1+\lambda_6^{-1} d\Z0.
\end{equation}
As for the attractor solutions, one reads $\varphi_1,~\varphi_2$
from~(\ref{twoscalars}), while $\varphi_3$ is given by
\begin{equation}\label{Solscalars}
\varphi_3 = \frac{2\nu}{\lambda_1}\ln{t} -\frac{1}{\lambda_6}\ln
\arrowvert{\varphi_3^{(0)}}\arrowvert
+\frac{\lambda_5}{\lambda_3\lambda_6} \ln \arrowvert
{\varphi_2^{(0)}}\arrowvert +
\frac{\lambda_3\lambda_4-\lambda_2\lambda_5}{\lambda_1\lambda_3\lambda_6}
\ln\arrowvert {\varphi_1^{(0)}}\arrowvert.
\end{equation}
The constants $\varphi_i^{(0)}$ are now given by
\begin{eqnarray}
\frac{\Lambda_1 \varphi_1^{(0)}}{3\gamma-1}&=&\frac{2}{\lambda_1
^2}\left(1-\frac{\lambda_2(\lambda_5^2+\lambda_6^2)
-\lambda_1\lambda_3\lambda_5}{\lambda_3\lambda_6^2}
\mu-\frac{\lambda_4\nu}{\lambda_3\lambda_6}\right), \nonumber
\\ \frac{\Lambda_2
\varphi_2^{(0)}}{3\gamma-1}&=&\frac{2}{\lambda_1\lambda_3}\left(\mu
-\frac{\lambda_5\nu}{\lambda_6}\right), \quad \frac{\Lambda_3
\varphi_3^{(0)}}{(3\gamma-1)} = \frac{2\nu}{\lambda_1\lambda_6}.
\end{eqnarray}
Notice that only the product $\Lambda_i \varphi_i^{(0)}$ is fixed
but not each term separately, so the increase of $\varphi_1^{(0)}$
to $\tilde{\varphi}_1^{(0)}$ can be absorbed by rescaling
$\Lambda_1\to
\Lambda_1\,(\tilde{\varphi}_1^{(0)}/\varphi_1^{(0)})$. The scalar
fields $\varphi_i$ transform under scale transformations in a
canonical way, i.e.  $\varphi_i\to \varphi_i/L$. 

\section{de Sitter vacuum and cosmic acceleration}\label{4dcosmology}

\subsection{Exponential potentials with more than one scalar}

Let us first discuss some general features of the cosmological
potential~(\ref{MainPoten1}) with $F^2=0$. The corresponding
solution will then exhibit a period of transient acceleration,
with the number of e-folds
\begin{equation}
N_e= \frac{n-1}{2(n-2)}\ln\frac{3-\sqrt{3}}{3+\sqrt{3}}
\frac{\sqrt{n+1}+\sqrt{n-1}}{\sqrt{n+1}-\sqrt{n-1}} +
\sqrt{\frac{n-1}{3}} \frac{1}{\sqrt{3(n-1)}-\sqrt{n+1}}\ln
\frac{\sqrt{3}+1}{\sqrt{3}-1},
\end{equation}
where $n\equiv p+2q+1$. Thus the number of e-folds increases only
marginally when the dimensions of the internal space are increased.
The late time behavior of the scale factor is
characterized by $a\propto t^\gamma$, where $\gamma=(n+1)/(n-1)$.

Let us consider some special cases, where one of the scalars
$\varphi_i$ takes a (nearly) constant value. In this rather
restricted case there can exist solutions with a large number of
e-folds. One such example is to consider the
potential~(\ref{MainPoten1}), with $\Lambda_2=0,\, p=4,\, q=1$ and
$\varphi_1=b_1=${const}. Of course this will constrain the
evolution of volume moduli and hence the 4d effective potential.
In this case, the scalar potential is given by
\begin{equation}\label{varphi1=0}
V={\rm e}^{-b_1}\left(\Lambda_1 {\rm e}^{-\sqrt{2}\varphi_2}+ F^2
{\rm e}^{-2b_1} {\rm
e}^{\varphi_2/\sqrt{2}-3\varphi_3/\sqrt{2}}\right).
\end{equation}
The explicit exact solution can be found in terms of a new
logarithmic time variable $\tau$, defined by $\tau=\int^t\dd \bar
t\,\exp\left[-\vph_2(\bar t)/\sqrt{2}\right]$. The explicit
solution is
\begin{eqnarray}
&&a=a_0 \left(\cosh\zeta\tau\right)^{\delta_-}
\left(\sinh\zeta\tau\right)^{\delta_+},\quad
b\Z1=\ln\frac{3\Lambda_1}{2V_0},\nonumber \\
&&\varphi_2 =\sqrt{\frac{18}{5}}
\ln\left(\left(\cosh\zeta\tau\right)^{\delta_-}
\left(\sinh\zeta\tau\right)^{-\delta_+}\right), \quad
\varphi_3=\varphi_2+ \frac{\sqrt{2}}{3}\ln\left(\frac{8}{9}
\frac{F^2 V_0^2}{\Lambda_1^3}\right),
\end{eqnarray}
where
\begin{equation}
\delta_{\pm}=\frac{\sqrt{6}}{3(\sqrt{6}\pm 1)},\quad
\zeta=\sqrt{\frac{5V_0}{8}},
\end{equation}
The total number of e-folds is given by
\begin{eqnarray}\label{efold-phi1=0}
N_e&=&\int H\dd t\nonumber \\
&=&\int_{\tau_0}^{\tau_-} \frac{a^\prime}{a}\dd\tau +
\int_{\tau_-}^{\tau_+} \frac{a^\prime}{a}\dd\tau.
\end{eqnarray}
The contribution of the first integral will be known once $\tau_0$
is chosen, while that of the second term depends on $\tau_+$. As
some representative values, with $V_0=1$, we find
$$
N_e=55,~ 60,~ 65, \cdots,
$$
respectively, when $\delta\varphi_2=41.91,\, 45.78,\, 49.65
\cdots$ (see Fig.~(3)). This gives only the minimum number of
e-folds, since there is an extra contribution from the first
integral in~(\ref{efold-phi1=0}) (i.e.  prior to the accelerated
epoch), which we have dropped here.

\begin{figure}[ht]
\vbox{\vskip-10.5pt\centerline{\scalebox{0.65}{\includegraphics{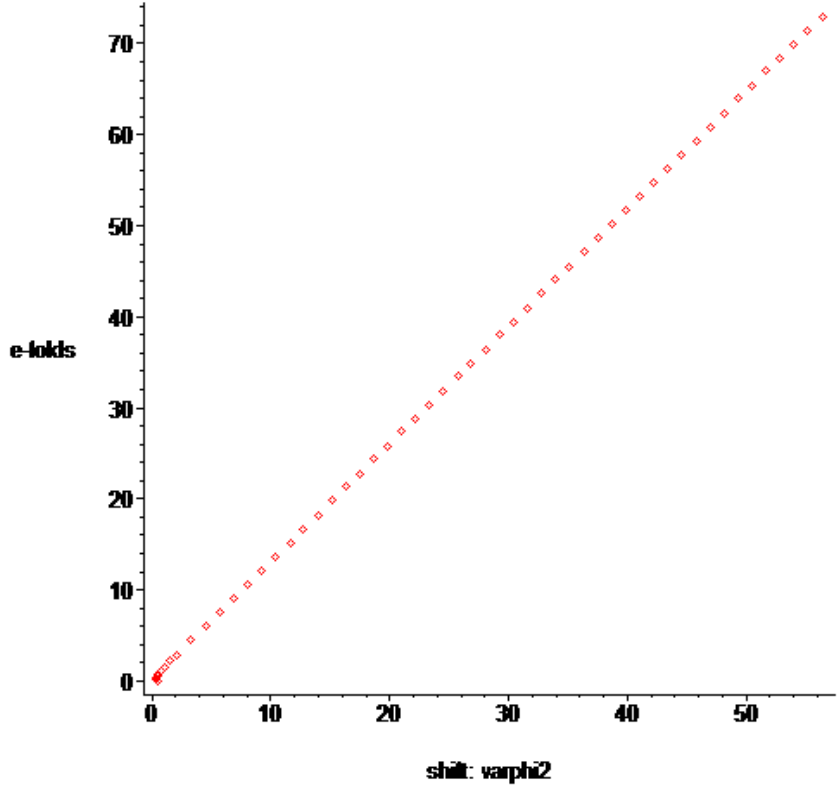}}}
\caption{\label{fig3}%
{\sl The number of e--folds $N$ vs scalar field $\varphi_2$, for
the potential~(\ref{varphi1=0}). }}}
\end{figure}

Let us consider some other possibilities, by allowing certain fine
tunings among the scalars $\varphi_i$ or the volume moduli
$\phi_i$. For the potential~(\ref{MainPoten1}), the late time
behavior of the scale factor may be characterized by $a(t)\propto
t^\gamma$, where
\begin{widetext}
\begin{equation}\label{GforGenPoten}
\gamma = \left\{\begin{array}{l}
\frac{q+1}{q}+\frac{2(p+2q+2)}{q^2(p+2q+3)} \qquad
(\varphi_2=\text{const},\;
\Lambda_2=0,\; \Lambda_1\ne0,\; F^2\ne0)\\
\frac{p(q+3)+q+1}{p+2q+3} \qquad \qquad (\varphi_1=\text{const},\;
\Lambda_2=0,\; \Lambda_1\ne0,\; F^2\ne0)\\
\frac{q}{q+1}+\frac{p+2q+2}{2(q+1)^2(p+2q+3)} \qquad
(\varphi_2=\text{const},\; \Lambda_1=0,\; \Lambda_2\ne 0,\; F^2\ne
0).
\end{array} \right.
\end{equation}
\end{widetext}
For instance, if $p=q=2$, the values of the expansion parameter
$\gamma$, in the above three cases, are $35/18$, $13/9$ and
$58/81$, respectively. In the first two cases, the amount of
inflation (or the number of e-folds) can be (arbitrarily) large
while it is small in the last case. In our model, when all scalar
fields $\varphi_i$ vary with time, then a positive potential
minimum representing the de Sitter phase of our universe can only
be metastable. The different choice in the potentials could lead
to the different asymptotic expansion of our universe.

\subsection{A model of eternal acceleration}

Let us extend the above discussion by specializing the case with
$p=4$ and $q=1$, but now we take the 4-space ${\cal M}^4$ as a
direct product between two $\MM2$,s of arbitrary curvatures. The
corresponding 11d metric {\it Ansatz} is
\begin{equation}\label{11dmetric}
\ds_{11}^2 = \e^{-2\Phi}\ds_4^2 + r_1^2 \e^{2\phi_1} \ds^2 (\MM
2_{\epsilon_1}) + r_2^2 \e^{2\phi_2}\ds^2 (\MM 2_{\epsilon_2})
+r_3^2 \e^{2\phi_3} ({dx}^2+{dy}^2) + r_4^2 \e^{2\phi_4}\left(dz +
\frac{f}{2} (x {dy}- y {dx}) \right)^2,
\end{equation}
where $\Phi=\phi_1+\phi_2+\phi_3+\phi_4/2$, and $\ds_4^2$ is the
standard FRW metric in the form~(\ref{FRWmetric}). Here we work in
units where the curvature radii $r_i$ are set to unity, as these
variables may often be absorbed in $\epsilon_i$ (i.e.
$\epsilon_i\to \epsilon_i/r_i^2$, where $\epsilon_i=0, \pm 1$)
and/or in the twist parameter $f$. In the gauge $\delta=0$, upon
reductions to four dimensions, the kinetic and potential terms are
given by
\begin{eqnarray}
K&=&2{\dot\phi_1}^2+2{\dot\phi_2}^2+2{\dot\phi_3}^2+
3{\dot\phi_1}^2/4+2\dot\phi_1\dot\phi_2+ 2\dot\phi_2\dot\phi_3  +
2\dot\phi_3\dot\phi_1+\dot\phi_1\dot\phi_4+ \dot\phi_2\dot\phi_4
+\dot\phi_3\dot\phi_4,\nonumber \\
V &=& -\epsilon_1 e^{-4\phi_1-2\phi_2-2\phi_3-\phi_4} -\epsilon_2
e^{-2\phi_1-4\phi_2-2\phi_3-\phi_4} + \frac{f^2}{4}
e^{-2\phi_1-2\phi_2-6\phi_3+ \phi_4}.
\end{eqnarray}
In terms of alternative canonically
normalized scalars, which may be defined by
\begin{eqnarray}
&&\varphi_1=\phi_1+\phi_2+2\phi_3+\phi_4/2, \quad
\varphi_2=\sqrt{3}\left(\phi_1+\frac{1}{3} \phi_2+\frac{1}{6}
\phi_4\right),
\nonumber \\
&&\varphi_3=\sqrt{\frac{8}{3}}\left(\phi_2+\frac{1}{8} \phi_4\right),
\quad \varphi_4=-\sqrt{\frac{9}{8}}\phi_4,
\end{eqnarray}
the corresponding field equations reduce to Eqs~(\ref{waveeq}),
(\ref{Fried1}) with $K=(1/2)\sum_i^4\dot{\varphi_i}^2$. The
resulting scalar potential is
\begin{equation}\label{11dPoten3}
V= -\epsilon_1 {\rm e}^{-\varphi_1-\sqrt{3}\varphi_2}-\epsilon_2
\e^{-\varphi_1 -\sqrt{\frac{1}{3}}\varphi_2
-\sqrt{\frac{8}{3}}\varphi_3} + \frac{f^2}{4} {\rm
e}^{-3\varphi_1+\sqrt{\frac{1}{3}}\varphi_2
+\sqrt{\frac{1}{6}}\varphi_3 -\sqrt{\frac{9}{2}}\varphi_4}.
\end{equation}
For the potential (\ref{11dPoten3}), there is a local
de Sitter minimum along $\varphi_2$, which is given by
\begin{equation}
\varphi_2^{(0)}=-\sqrt{3}\varphi_1-2\sqrt{2}\varphi_3 -
\frac{\sqrt{3}}{2}\ln \frac{\pm \sqrt{A^2- 3 B}- A}{6},
\end{equation}
where
\begin{equation}
A\equiv \frac{\epsilon_2}{\epsilon_1} {\rm
e}^{-2\varphi_1-2\sqrt{6}\varphi_3}, \quad B\equiv
\frac{{f}^2}{\epsilon_1} {\rm
e}^{-6\varphi_1-\frac{5\sqrt{6}}{2}\varphi_3-
\frac{3}{\sqrt{2}}\varphi_4}.\nonumber \\
\end{equation}
Of course, we are allowing here for the possibility that $A<0$,
$B>0$ (i.e.  $\epsilon_1=+1, \epsilon_2=-1$), in which case there
can arise two local minima along $\varphi_2$. With $A<0$ (or
$A>0$) and $B<0$, there is only one local (de Sitter) minimum. The
solution with $A>0$, $B>0$ (i.e.  $\epsilon_1=\epsilon_2=+1$) is,
however, unstable.

Let us consider a special case where $\varphi_1=\mbox{const}\equiv
b\Z0$. An explicit exact solution can be found in terms of a
logarithmic time variable $\tau$, defined by
$\tau=\int^t\dd\bar{t} \exp[-\sqrt{\frac{3}{4}}\,
\varphi_2(\bar{t})]$. The solution is
\begin{eqnarray}
&&a=a\Z0 \left(\cosh\zeta\tau\right)^{\delta_-}
\left(\sinh\zeta\tau\right)^{\delta_-}, \quad \varphi_1=b\Z0,\quad
\varphi_2=\sqrt{2} \ln\left[(\cosh\zeta\tau)^{\delta_-}
(\sinh\zeta\tau)^{-\delta_+}\right],
\nonumber \\
&&\varphi_3=\frac{\varphi_2}{\sqrt{2}} + \sqrt{\frac{3}{32}}
\ln\left(\frac{3\epsilon_2^2}{2\epsilon_1^2}\right), \quad
\varphi_4=\sqrt{\frac{3}{2}}\varphi_2+\frac{1}{3\sqrt{2}}
\left(\ln\frac{3f^4}{32\epsilon_1^2}-4b\Z0\right),
\end{eqnarray}
where
\begin{equation}
\delta_{\pm}=\frac{\sqrt{6}}{3(\sqrt{6}\pm 1)}, \quad
\zeta=\sqrt{\frac{-5\,\epsilon_1}{8}}\sqrt{1+\sqrt{\frac{8}{3}}}\,
\e^{-b\Z0/2}.
\end{equation}
The late time behavior of the solution is characterized by
\begin{eqnarray}
&&a(t)\propto t^2, \quad \varphi_1=b_0,
\quad \varphi_2= \sqrt{\frac{4}{3}}\ln{t}+\mbox{const}, \nonumber \\
&& \varphi_3= \sqrt{\frac{2}{3}}\ln{t}+\mbox{const}, \quad
\varphi_4=\sqrt{2} \ln{t}+\mbox{const}.
\end{eqnarray}
In terms of the original volume scalars $\phi_i$, we find
\begin{equation}
\e^{\phi_1}, \e^{\phi_2}\propto t^{2/3},\quad
\e^{\phi_3}\propto t^{-1/3}, \quad \e^{\phi_4}\propto
t^{-4/3}.
\end{equation}
In this example the space $(T^2\ltimes S^1)$ shrinks with time,
though with different time-varying factors, while the 4-space
${\cal M}^4$ expands.

\section{More than one geometric twist}

\subsection{string/M theory with double twist}

Consider an 11d metric {\it ansatz} such that
\begin{equation}\label{11Ddecom}
\Sigma_7= (\MM 2_{\epsilon_1}\ltimes S^1)\times (\MM
2_{\epsilon_2}\ltimes S^1)\times S^1.
\end{equation}
We assign to each factor spaces the different time-varying scales:
$e^{\phi_1}, e^{\phi_2}, \cdots, e^{\phi_5}$, where, in general,
$\phi_1\neq \phi_2$, $\phi_3\neq \phi_4$. We also introduce the
two non-trivial twist parameters, $f_1$ and $f_2$, and choose the
Einstein conformal frame by setting
$\Phi=\phi_1+\phi_2/2+\phi_3+\phi_4/2+\phi_5/2$.

We will begin by presenting a special exact solution where
$\epsilon_i=0$ (so $\MM 2 \to \TI^2$) and each 2-space $\TI^2$ has
the same time-varying volume. In this case, in the gauge
$\delta=3$, the solution will be qualitatively similar to that
with a single twist, namely,
\begin{eqnarray}\label{Restricted1}
&&a= a_0 \cosh(\chi {u})\, \e^{-\frac{1}{2}(8c+c_2)u},\nonumber \\
&&\phi_1=\ln \cosh\chi {u} -\frac{1}{2}(6c+c_2)u
=\phi_3,\nonumber \\
&&\phi_2=-\phi_1+c_1 u+\ln\frac{f_2}{f_1}
+\phi_0=\phi_4+\ln\frac{f_2}{f_1}, \nonumber\\
&&\phi_5=c_2 u, \quad f_1 f_2=2\chi^2/a_0^6,
\end{eqnarray}
where
\begin{equation}
\chi=\frac{1}{2} \sqrt{48c^2+12 c c_2-8 c_1^2-2 c_2^2-4 c_1 c_2}.
\end{equation}
This solution will of course exhibit only a period of transient
acceleration, with number of e-folds ${N}_e={\cal O}(1)$.
However, there can arise new solutions with a larger number of
e-folds, especially, with a non-zero curvature, $\epsilon_i\neq
0$.

Let us consider another canonical example, but with six extra
dimensions, i.e.  $\Sigma_6=(\TI^2\ltimes S^1)\times (\TI^2\ltimes
S^1)$. In this case, the explicit solution may be given by
\begin{eqnarray}\label{n=6Sol}
&& a(u)= a\Z0 \cosh(2 c {u})\, \e^{{4} c u/\sqrt{3}},
\nonumber \\
&& \phi_1 = \phi_3=\sqrt{3} c {u} +\ln \cosh (2 c{u}), \nonumber \\
&&\phi_2=-\phi_1+\ln \arrowvert \frac{f_2}{f_1}\arrowvert
+\phi_0=\phi_4+\ln \arrowvert \frac{f_2}{f_1}\arrowvert,
\end{eqnarray}
subject to the constraint
\begin{equation}
\quad f_1 f_2=\frac{8c^2}{a\Z0^6},
\end{equation}
where $a\Z0$, $c$ and $\phi\Z0$ are integration constants. This
solution can be obtained also from~(\ref{Restricted1}) after the
substitutions: $c\to - c/\sqrt{3}$, $c_1=0$ and also $c_2=0$, as
there is no $\phi_5$.

\subsection{Some special solutions}

Let us consider that the internal 7-space is split as
in~(\ref{11Ddecom}). But now we take $\epsilon_i\ne 0$. Upon
reductions to four dimensions, we find that the scalar potential
due to the non-trivial geometric twists, $f_i>0$, is given by
\begin{equation}
V (f_i, \varphi_i) = \frac{f_2^2}{4}
e^{-\sqrt{\frac{2}{3}}\varphi_1-\sqrt{\frac{1}{3}}\varphi_2
-\sqrt{\frac{25}{3}}\varphi_3+\sqrt{\frac{1}{6}}\varphi_4
-\sqrt{\frac{9}{2}}\varphi_5} +\frac{f_1^2}{4}
\,e^{\sqrt{\frac{2}{3}}\varphi_1-\sqrt{\frac{4}{3}}\varphi_2
-\sqrt{\frac{4}{3}}\varphi_3 -\sqrt{\frac{32}{3}}\varphi_4},
\end{equation}
while, the curvature contribution to a scalar potential is given
by
\begin{equation}
V (\epsilon_i, \varphi_i)= - {\epsilon_1}\,
e^{-\sqrt{\frac{2}{3}}\varphi_1-\sqrt{\frac{1}{3}}\varphi_2
-\sqrt{\frac{1}{3}}\varphi_3-\sqrt{\frac{8}{3}}\varphi_4}
-{\epsilon_2}
\,e^{-\sqrt{\frac{2}{3}}\varphi_1-\sqrt{\frac{1}{3}}\varphi_2
-\sqrt{3}\varphi_3},
\end{equation}
where $\varphi_i$ are canonically normalized four-dimensional
scalars, defined in the following combinations:
\begin{eqnarray}
&&\varphi_1=\sqrt{\frac{1}{6}}(2\phi_1+3\phi_2+2\phi_3+\phi_4+\phi_5),
\quad
\varphi_2=\sqrt{\frac{1}{12}}(2\phi_1+2\phi_3+4\phi_4+\phi_5),
\nonumber \\
&&\varphi_3=\sqrt{\frac{1}{3}}(\phi_1+3\phi_3+\phi_5/2),\quad
\varphi_4=\sqrt{\frac{1}{24}}(8\phi_1+\phi_5),\quad
\varphi_5=-\sqrt{\frac{9}{8}}\phi_5.
\end{eqnarray}
The total scalar potential is then given by $V=V(f_i, \varphi_i) +
V(\epsilon_i, \varphi_i)$.

One may find various exact solutions by specializing the potential
$V$ to some particular models, such as\
\begin{eqnarray}
&& \varphi_1,\,\varphi_2=\mbox{const} \Rightarrow
\phi_2=\phi_4+\mbox{const},\nonumber \\
&&\varphi_2,\, \varphi_3=\mbox{const}\Rightarrow
\phi_3 =\phi_4+\mbox{const}, \nonumber \\
&&\varphi_1,\,\varphi_3=\mbox{const} \Rightarrow
3\phi_2+\phi_4-4\phi_3=\mbox{const}.\nonumber
\end{eqnarray}
Let us consider the first case, where each 2-space $\MM 2$ will
have the same time-varying volume, by further setting $f_1=0$,
$\epsilon_1=+1$ and $\epsilon_2=-1$, but $f\Z2>0$. In terms of a
logarithmic time variable $\tau=\int^t\dd\bar{t} \exp[-\sqrt{3}\,
\varphi_3(\bar{t})/2]$ or $\dd \tau=\e^{-\sqrt{3}\varphi_3/2} \dd
t$, the explicit solution is found to be
\begin{eqnarray}
&&a=a\Z0 \left(\cosh\zeta\tau\right)^{\delta_-}
\left(\sinh\zeta\tau\right)^{\delta_-},\nonumber\\
&& \varphi_1=b\Z1=\mbox{const},\quad \varphi_2=b\Z2=\mbox{const},
\nonumber \\
&&\varphi_3=\sqrt{\frac{18}{5}}
\ln\left[(\cosh\zeta\tau)^{\delta_-}
(\sinh\zeta\tau)^{-\delta_+}\right]+\varphi_0,
\nonumber \\
&&\varphi_4=\frac{\varphi_3}{\sqrt{2}} -
\frac{\sqrt{3}}{4\sqrt{2}}\left(\ln{3}-\ln{2}\right),
\nonumber \\
&&\varphi_5=-\frac{\varphi_3}{\sqrt{6}}-
\frac{1}{3\sqrt{2}}
\left(\ln{3}+\ln{2}-4\ln{f_2}\right),\nonumber \\
\end{eqnarray}
where
\begin{equation}
\delta_{\pm}=\frac{\sqrt{10}}{3(\sqrt{10}\pm \sqrt{3})}, \quad
\zeta=\sqrt{\frac{21(4-\sqrt{6})}{160}}
\e^{-(\sqrt{2}b_1+b_2)/\sqrt{12}}.
\end{equation}
The late time behavior of the solution may be characterized by
\begin{eqnarray}
&&a(t)\propto t^{10/9}, \quad \varphi_3=
\sqrt{\frac{4}{3}}\ln{t}+\mbox{const},
\nonumber\\
&& \varphi_4= \sqrt{\frac{2}{3}}\ln{t}+\mbox{const}, \quad
\varphi_5=\sqrt{\frac{2}{9}} \ln{t}+\mbox{const}.
\end{eqnarray}
In terms of the original scalars $\phi_i$, we find
\begin{equation}
\e^{\phi_1}, \e^{\phi_3}, \e^{\phi_5}
\propto t^{4/9},\quad
\e^{\phi_2}, \e^{\phi_4}
\propto t^{-5/9}.
\end{equation}
In all these solutions the physical three space dimensions expand
faster than the remaining ones. In the zero-twist case, i.e.
$f_1=0,~f_2=0$, we find that, when $\epsilon_i=-1$, the asymptotic
behavior of the scale factor is $a(t)\propto t^{2/3}$. This
behavior, however, can be different in the presence matter fields,
like, dust ($\omega=1$) and radiation ($\omega=4/3$).

\subsection{Slowly rolling moduli}

In this subsection we show that there exist cosmological
compactifications of a $4+n$ dimensional Einstein gravity on
twisted spaces of time-dependent metric where 3-space dimensions
expand much faster than the remaining $n$. To quantify this, let
us consider a decomposition
$$
\Sigma =(\TI^2\times \TI^2)\ltimes S^1 \times (\TI^2\times\TI^2)
\ltimes S^1, $$ by assigning different time-varying scale factors
${\rm e}^{\phi_1},\cdots, {\rm e}^{\phi_4}$, respectively. We also
assign the two non-zero twist parameters $f_1$ and $f_2$, and
choose the 4d Einstein conformal-frame by setting
$\Phi=2\phi_1+\phi_2/2+2\phi_3+\phi_4/2$. While it is not clear if
this model is phenomenologically viable within the string/M-theory
context, mainly because the number of extra dimensions $n$ is
$>7$, it nonetheless provides an interesting example where at late
times the size of extra dimensions can be much smaller compared to
the size of the physical universe.

Let us begin by presenting a special exact solution where each
factor space $\TI^2$ has the same volume. The solution is most
readily written down in the gauge $\delta=3$, which is given by
\begin{eqnarray}
&& a= a\Z0 \,{\rm e}^{2\left(d^2/c+c\right) u}\,
\e^{\frac{2}{(24c)^2}\, \e^{24 c {u}}}, \nonumber \\
&&\phi_1= \phi_3= (d^2/c-c) {u}+\frac{1}{(24c)^2}\,
\e^{24 c{u}},\nonumber\\
&&\phi_2=-\phi_1+ d u+ \ln\frac{f_2}{f_1}+\phi_0
=\phi_4+\ln \frac{f_2}{f_1},
\end{eqnarray}
subject to the constraint
\begin{equation}
f\Z1 f\Z2 = 2/a\Z0^6.
\end{equation}
where $a_0, c$ and $d$ are integration constants.
This solution exhibit a period of
transient acceleration when $ c^2> 2 d^2$.
Acceleration occurs on the interval $u_+<u < u_-$, where
\begin{equation}
u_{\pm} =\frac{1}{24c} \ln \left[ 24(2c^2-d^2)\pm 24\sqrt{3c^2(c^2-2
d^2)} \right].
\end{equation}
One also notes that
\begin{equation}
{a}= a_0 \,
\e^{\left(3c+\frac{d^2}{c}\right)u+\left(\frac{1}{24c}\right)^2
\e^{24 c{u}}}\, \e^{\phi_1}.
\end{equation}
Clearly, our solutions contain new parameters, the time
$u=\bar{u}<0$ at which all space dimensions have comparable size,
$a\sim a_0 \e^{\phi_i}$. However, at late times ($u\to +\infty$),
the ratio between the two scale factors can be such that the size
of the internal space is much smaller compared to the scale factor
of physical 3-space.

A solution qualitatively similar to the above one arises also for
the following decomposition
$$
\Sigma_n=(\TI^2\ltimes S^1)\times (\TI^2\ltimes S^1)\times
(\TI^2\ltimes S^1),$$ with the time-varying scale factors
$\e^{\phi_1}, \cdots, \e^{\phi_6}$, respectively. We have been
able to find the explicit solution only in the case when each
2-space $\TI^2$ has the same scale factor, which is given by
\begin{eqnarray}
&& a= a_0
\exp\left({\left(\frac{\alpha}{12}+\frac{11}{3}\frac{d^2}{\alpha}\right)
u}\right) \, \exp\left({\frac{3}{2} \frac{1}{\alpha^2}
\e^{\alpha u}}\right), \nonumber \\
&& \phi_1=\phi_3=\phi_5=\frac{1}{\alpha^2} \e^{\alpha u}
-\frac{(9c^2-15 d^2) u}{2 \alpha}, \nonumber \\
&&\phi_2=-\phi_1- d\, u+\ln  \frac{f_2 f_3}{2},
\nonumber \\
&&\phi_4=-\phi_3 - d\, u+\ln \frac{f_1 f_3}{2},
\nonumber \\
&& \phi_6=-\phi_5 - d\, u+\ln \frac{f_1 f_2}{2},
\end{eqnarray}
where $\alpha\equiv 9c+d$,with $c$ and $d$ being the integration
constants. It is possible that the exact solutions that we have
written above correspond to some special cases, where each 2-space
$T^2$ has the same volume, while, in general, each space in the
product spaces can have a different time-varying scale factor. In
any case, these examples clearly show that a universe with $3+n$
space dimensions, even of comparable size at early time, could
evolve to a universe in which the 3 space dimensions become much
larger than the remaining $n$ dimensions at late times.

\subsection{A scalar potential of appropriate slope}

It is well appreciated that a single scalar in exponential form
$V= V_0 e^{-\lambda\varphi}$, with the slope
$\lambda\leq \sqrt{2}$, can reasonably explain the cosmological
inflation with the number of e-folds $N_e\geq 11.68$ (cf. the
discussion in Section III), see,
e.g.,~\cite{Kallosh:2002gf,Kehagias:2004bd}.
Whereas such models of string/M theory
origin may be quite interesting,
no explicit model of this type has been constructed so far from
compactification (or higher dimensional gravity).
Here we give a simple example where this may
be accomplished.

Consider that the internal 7-space $\Sigma_7$ is split as
$$ \Sigma_7 = {\cal M}^{6}_{\epsilon_1}\ltimes S^1 $$
with the scale factors $r_1 e^{\phi_1}$ and $ r_2 e^{\phi_2}$,
respectively. The parameter $\varpi$ (appeared
in~(\ref{Twisted1})) may be realized now as $f$ (which is a twist
parameter) times one-form on $\MM 6$. A simplification occurs
when $\MM 6$ is split as $\MM
2\times \MM 2\times \MM 2$. One also chooses the Einstein
conformal frame by setting $\Phi=3\phi_1+\phi_2/2$. Let us choose
the gauge $\delta=0$. The kinetic and potential terms are then
given by
\begin{eqnarray}
&& K=\frac{1}{2} \left(24\dot{\phi_1}^2+\frac{3}{2}\dot{\phi_2}^2
+6\dot{\phi_1}\dot{\phi_2}\right), \nonumber \\
&& V= -\frac{3\epsilon_1}{r_1^2}\,\e^{-8\phi_1-\phi_2}
+\frac{f^2}{4}\frac{r_1^2}{r_2^4}\,\e^{-10\phi_1+\phi_2}.
\end{eqnarray}
In terms of alternative
canonically normalized scalars $\varphi_i$, which may be defined by
$$\varphi_1= \sqrt{18}\,\phi_1, \qquad
\varphi_2=\pm \sqrt{6}\left(\phi_1+\phi_2/2\right),$$
the kinetic term is
$K=\frac{1}{2}(\dot{\varphi_1}^2+\dot{\varphi_2}^2)$. The
resulting scalar potential may be given by
\begin{equation}\label{appro1}
V= -\frac{3\epsilon_1}{r_1^2}\,
e^{-\sqrt{2} \varphi_1-\sqrt{2/3}\,\varphi_2}
+\frac{f^2}{4}\frac{r_1^2}{r_2^4}\,
e^{-\sqrt{6} \varphi_1+\sqrt{2/3}\,\varphi_2}.
\end{equation}
In the case $\epsilon_1=+1$, we demand $f>\sqrt{12} (r_2/r_1)^2$,
so that $V>0$ when $\varphi_1 \simeq \varphi_2\simeq 0$, but there
is no such restriction for $\epsilon_1=0$ and $\epsilon_1=-1$. The
potential~(\ref{appro1}) has a positive definite minimum along
$\varphi_1$ (if $\epsilon_1=+1$) or along $\varphi_2$ (if
$\epsilon_1=-1$). If any such minimum represents the present de
Sitter phase of our universe, then it can be only metastable. In
this case both the fields $\varphi_1$ and $\varphi_2$ vary with
time. However, a physically different (and perhaps more
interesting) situation arises when $\varphi_2$ takes a (nearly)
constant value (i.e. $\phi_1 \sim -\phi_2/2+\mbox{const} $), in
which case when the size of $S^1$ expands, the size of ${\cal
M}^6$ shrinks, providing a ``4+1+compact space" type background.

\subsection{A canonical example}

Finally, as the most canonical example, assume that in the
four-dimensional effective theory the scalar potential is
parameterized by $V=V_0 e^{-\lambda\varphi}$. One may think of
this potential to arise from a compactification of (classical)
supergravity on spaces with a time-dependent
metric~\cite{CHNW03b,Townsend03b}, or from a standard flux
compactification in string theory. The scalar wave equation is
then
\begin{equation}
\ddot{\varphi}+3 H \dot{\varphi} -\lambda V\Z0
\e^{-\lambda\varphi}=0.
\end{equation}
Let us also include the possibility of a stress
tensor for which the mass-energy density evolves as
\begin{equation} \dot{\rho_\omega} =-3 H (P_\omega+\rho_\omega),
\end{equation}
where $P_\omega\equiv
(\omega-1)\rho_\omega$.
In general, the energy component $\rho_\omega$ is not explicitly
coupled to the scalar field, but gravitationally coupled via the
modified Friedmann constraint for the Hubble expansion:
\begin{equation}
H^2=\frac{1}{3}\left(\frac{1}{2}\dot{\varphi}^2+V_0 e^{-\lambda
\varphi} +\frac{c_0}{a^{3\omega}}\right).
\end{equation}
Clearly the spatial curvature term $k/a^2$ acts as a fluid
density with $\omega=2/3$; in this context, a physically relevant
case is $k=-1$ (i.e. $c_0> 0$). The general solution to the
above set of equations may be written explicitly in terms of
cosmic time, $t$, when $\omega=2$ (stiff matter), while in other
cases, like $\omega=1$ (dust) or $\omega=4/3$ (radiation), it
might be necessary to adopt some other time-coordinates. Fixed
point (asymptotic) solutions of the evolution equations with
$\omega=2/3$ have recently appeared in Ref.~\cite{Giddings04a}.

With $\omega=2$, the explicit exact solution is given by
\begin{eqnarray}
&&a= a_0
\left(t^{6/\lambda^2}+\frac{t}{\lambda^2}\right)^{1/3},
\nonumber \\
&&\varphi=\frac{2}{\lambda}\ln t +\ln
\left|\frac{\lambda^2}{6-\lambda^2}\frac{V_0}{2}
\left(t^{(\lambda^2-6)/\lambda^2}+ \lambda^2 \right)\right|,
\nonumber \\
&&\rho= \frac{(\lambda^2-6)^2(2\lambda^2-3)}{3\lambda^6} \left(t+
\lambda^2 t^{6/\lambda^2}\right)^{-2},
\end{eqnarray}
where $\lambda^2\neq 6$. With $\lambda^2 = 2$, one would have a
linear regime $a\sim t$ asymptotically. In the $\lambda^2\leq 2$
case, there is no cosmological event horizon, and so it does not
suffer from the problems with the existence of an event horizon
discussed, e.g., in~\cite{Susskind01a}. The $\omega=2$ case
corresponds to a stiff matter with the equation of state
$P_\omega=\rho_\omega$. In such a medium, the velocity of sound
approaches to that of light and hence the cosmology with
$\omega=2$ may differ considerably from the usually contemplated
scenarios with $\omega=1$. We will return to the $\omega\neq 2$
case in a future publication.

\medskip
\section{Discussion and conclusion}

In Ref.~\cite{TW}, Townsend and Wohlfarth studied Kasner type
solutions of a pure gravitational theory in $4+n$ dimensions which
avoids the no-go theorem of~\cite{nogo1} by introducing negatively
curved (hyperbolic) extra dimensions with a time-dependent metric.
Here we have demonstrated that the same no-go theorem can be
circumvented by compactifications on {\it Ricci--flat} internal
spaces. This is possible through the introduction of one or more
geometric twists in the internal (product) space. That is, a
cosmological compactification of classical supergravities on {\it
Ricci-flat} spaces with a non-trivial twist can naturally give
rise to an expanding four-dimensional FRW universe undergoing a
period of transient acceleration in 4d Einstein frame, {\it
albeit} with e-folding of order unity.

There was no way in the TW approach to account for a preferential
compactification of a certain number of extra dimensions. In view of this
somewhat discouraging result, one might welcome a modification of
this system in which some of the extra dimensions shrink
with time and/or the runway potential is not too steep.

We have explained how the scalar potentials arising from
compactifications on maximally symmetric spaces are modified when
there exist one or more non-trivial twists. The combined effects
of the non-trivial curved spaces and the geometric twists could
result in a sum of exponential potentials in the effective 4d
theory which involves many fields and cross-coupling terms. Such
potentials are more likely to have a locally stable minimum with a
positive vacuum energy. By expressing the field equations in terms
of canonically normalized four--dimensional scalars, we have
obtained a general class of exact cosmological solutions for
multi--scalar fields with exponential potentials. We also
investigated the case where some of the scalar fields take
(nearly) constant values (which will then constrain the evolution
of volume moduli and hence the runway potential in the higher 10
or 11 dimensions) and demonstrated that a number of novel features
emerged in this scenario, including the possibility of increasing
the rate of expansion (or the number of e-folds) when there
existed a mixture of positive and negative slopes in the
potential. The basic characteristics of the models
(compactifications) studied in this paper are summarized in
Table $I$.

\begin{table}[tp!]
{TABLE I: } {A summary of compactifications, by internal space
$\Sigma_n ~(n=p+2q+1)$, showing the basic characteristics of the
models. The 2-dimensional space ${\cal M}^2$ is either $S^2$ or
$\HI^2$.}
\begin{tabular*}{\textwidth}{@{\extracolsep{\fill}} cccccc}\hline\hline
Internal space & Number of twists & Fine-tuning & Compactifying
space & Decompactifying space & Acceleration \\
\hline $\TI^p\times (\TI^2 \times \cdots \times \TI^2) \times
\RI^1$ & $ 0 $ & yes/no &
\multicolumn{2}{c}{different possibilities.}  & no \\
\hline $\TI^p\times \underbrace{\TI^2 \times \cdots \times
\TI^2}_{}\ltimes S^1$ & $q$ & no & $ \TI^2~(S^1)$ & $S^1~ (\TI^2)$
& transient
\\ \hline $\HI^p\times \underbrace{\TI^2\times \cdots\times \TI^2}_{}
\ltimes S^1$ & $q$ & no & $\TI^2~\text{or}~S^1$~ (or both) &
$\HI^p$ & transient
\\ \hline $\TI^p\times \underbrace{\HI^2\times \cdots\times \HI^2}_{}
\ltimes S^1$ & $ q $ & no &   \multicolumn{2}{c}{different
possibilities.} & transient \\ \hline $\HI^2\times {\cal
M}_{\epsilon_2}^2 \times (\TI^2\ltimes S^1)$ & $ 1 $ & yes &
$\TI^2$, $S^1$ & $\HI^2\times {\cal M}_{\epsilon_2}^2$ & eternal \\
\hline $\HI^4\times (\TI^2\ltimes S^1)$ & $ 1 $ & yes & $\TI^2$,
$S^1$
& $\HI^4$ & eternal \\
\hline $\HI^2\times \underbrace{\TI^2\times \TI^2}_{}\ltimes S^1$
& $ 2 $ & yes
& $\TI^2$, $S^1$ & $\HI^2$ & eternal \\
\hline $(\TI^2\ltimes S^1)\times (\TI^2\ltimes S^1)$ & $ 2 $ & no
& $\TI^2$ & $S^1$ & transient \\ \hline $ \underbrace{\TI^2 \times
\TI^2}_{} \ltimes S^1 \times \underbrace{\TI^2\times
\TI^2}_{}\ltimes S^1$ & $ 2, 2 $ & no & $\TI^2\times \TI^2 ~(S^1)$
& $S^1 ~ (\TI^2\times \TI^2)$ & transient \\
\hline $ (\TI^2\ltimes S^1) \times (\TI^2\ltimes S^1) \times
(\TI^2\times S^1)$ & $ 1, 3 $ & no & $\TI^2 ~(S^1)$
& $S^1 ~ (\TI^2)$ & transient \\
\hline $ ({\cal M}_{\epsilon_1}^2\ltimes S^1)\times ({\cal
M}^2_{\epsilon_2} \ltimes S^1) $ & 2& no/yes &
\multicolumn{2}{c}{different
possibilities.} & transient/eternal \\
\hline\hline
  \end{tabular*} 
\end{table}\label{table1}

For all cosmological solutions arising from compactifications on
spaces with a time-dependent metric the modulus field must give
rise to a runaway potential, if it is to allow an expanding
four-dimensional FRW universe undergoing a period of acceleration.
This follows from just accepting the fact that in an effective
four-dimensional cosmology these scalars can be time-dependent.
The rolling of the moduli can be minimized by introducing one or
more geometric twists along the extra space along with curved
internal spaces. So the issue should not be that one needs to find
a mechanism to stabilize the moduli completely, but rather that to
make any cosmological model arising from compactification
phenomenologically viable, it is required that the extra
dimensions become unobservably small at late times by contracting
or expanding at a much slower rate than the physically observable
dimensions.

The universe after inflation is dominated by matter fields (radiation
and dust). In higher-dimensional supergravity theories,
supersymmetry requires the presence of specific matter fields,
which may hold the clue as to why precisely three space
dimensions stay large up to the present epoch. In order to fully account
for the fate of the extra dimensions, it may be necessary to include
excitations of the Fermi fields in the cosmological solutions. If
so, our results are presumably restricted to an era close to the
dimensional transition, where the physical three space-like
dimensions become distinguished from the remaining 6 or 7.

In this paper we have studied the possibility of generating
inflation from higher dimensional gravity on product spaces, by
introducing only non--trivial curved spaces and/or certain
``twists'' in the geometry. It would be worthwhile to investigate
the effects of non-zero (electric) form-fields in spacetime
dependent (warped) compactification by turning on magnetic field
strength (or equivalently a geometric twist) and to study their
effects on the evolution of the 4d spacetime. Some earlier
papers~\cite{Kaloper:1999a,Pope:1996a}, involving a discussion of
the cosmological advantages of string/M theory compactifications
with twisted internal spaces and fluxes, may be of interest in
this context.

The discussion in \cite{Barreiro:1999zs} shows that the properties
of the attractor solutions of exponential potentials (with two
scalars) may lead to model of quintessence with currently
observationally favored equations of state, i.e.  $w\sim -1$. It
may be worthwhile to redo this analysis by considering
multi--scalar exponential potentials with the slopes predicted by
our specific examples. One possibility is that desirable slopes
can be obtained from the assisted behavior when one or more of the
fields take a relatively constant value for a sufficiently long
time during inflation. An obvious advantage of having multiple
exponential potentials is that there can exist more than one stage
of inflation, some of which happens in a vicinity of different
minimum of the effective potential.

\medskip

{\bf Acknowledgement}\ This work was supported in part by the
Mardsen fund of the Royal Society of New Zealand. I.P.N. is
grateful to Pei-Ming Ho for the collaboration in the course of
which he learned some of the ideas presented here and to Ed
Copeland for providing some useful insights on the problem.

\end{document}